\documentclass[twocolumn]{aastex631}  
\usepackage{natbib}
\usepackage{booktabs}
\usepackage{amsmath}
\usepackage[figuresright]{rotating}
\usepackage{url}
\usepackage{booktabs}

\begin{document}
	
\title{Multiwavelength Polarization Observations of Mrk 501}
	
\correspondingauthor{Jin Zhang, Xiang-Gao Wang, Kishore C. Patra}
\email{j.zhang@bit.edu.cn; wangxg@gxu.edu.cn; kcpatra@berkeley.edu}

\author{Xin-Ke Hu}
\affiliation{Guangxi Key Laboratory for Relativistic Astrophysics, School of Physical Science and Technology, Guangxi University, Nanning 530004, People's Republic of China}

\author{Yu-Wei Yu}
\affiliation{School of Physics, Beijing Institute of Technology, Beijing 100081, People's Republic of China}

\author{Jin Zhang\dag}
\affiliation{School of Physics, Beijing Institute of Technology, Beijing 100081, People's Republic of China}

\author{Xiang-Gao Wang\dag}
\affiliation{Guangxi Key Laboratory for Relativistic Astrophysics, School of Physical Science and Technology, Guangxi University, Nanning 530004, People's Republic of China}

\author{Kishore C. Patra\dag}
\affiliation{Department of Astronomy, University of California, Berkeley, CA 94720-3411, USA}

\author{Thomas G. Brink}
\affiliation{Department of Astronomy, University of California, Berkeley, CA 94720-3411, USA}

\author{Wei-Kang Zheng}
\affiliation{Department of Astronomy, University of California, Berkeley, CA 94720-3411, USA}

\author{Qi Wang}
\affiliation{Guangxi Key Laboratory for Relativistic Astrophysics, School of Physical Science and Technology, Guangxi University, Nanning 530004, People's Republic of China}

\author{De-Feng Kong}
\affiliation{Guangxi Key Laboratory for Relativistic Astrophysics, School of Physical Science and Technology, Guangxi University, Nanning 530004, People's Republic of China}

\author{Liang-Jun Chen}
\affiliation{Guangxi Key Laboratory for Relativistic Astrophysics, School of Physical Science and Technology, Guangxi University, Nanning 530004, People's Republic of China}

\author{Ji-Wang Zhou}
\affiliation{Guangxi Key Laboratory for Relativistic Astrophysics, School of Physical Science and Technology, Guangxi University, Nanning 530004, People's Republic of China}

\author{Jia-Xin Cao}
\affiliation{Guangxi Key Laboratory for Relativistic Astrophysics, School of Physical Science and Technology, Guangxi University, Nanning 530004, People's Republic of China}

\author{Ming-Xuan Lu}
\affiliation{Guangxi Key Laboratory for Relativistic Astrophysics, School of Physical Science and Technology, Guangxi University, Nanning 530004, People's Republic of China}

\author{Zi-Min Zhou}
\affiliation{Guangxi Key Laboratory for Relativistic Astrophysics, School of Physical Science and Technology, Guangxi University, Nanning 530004, People's Republic of China}

\author{Yi-Ning Wei}
\affiliation{Guangxi Key Laboratory for Relativistic Astrophysics, School of Physical Science and Technology, Guangxi University, Nanning 530004, People's Republic of China}

\author{Xin-Bo Huang}
\affiliation{Guangxi Key Laboratory for Relativistic Astrophysics, School of Physical Science and Technology, Guangxi University, Nanning 530004, People's Republic of China}

\author{Xing-Lin Li}
\affiliation{Guangxi Key Laboratory for Relativistic Astrophysics, School of Physical Science and Technology, Guangxi University, Nanning 530004, People's Republic of China}

\author{Hao Lou}
\affiliation{Guangxi Key Laboratory for Relativistic Astrophysics, School of Physical Science and Technology, Guangxi University, Nanning 530004, People's Republic of China}

\author{Ji-Rong Mao}
\affiliation{Yunnan Observatories, Chinese Academy of Sciences, 396 Yangfangwang, Guandu District, Kunming, 650216, People’s Republic of China}
\affiliation{Center for Astronomical Mega-Science, Chinese Academy of Sciences, 20A Datun Road, Chaoyang District, Beijing, 100012, People’s Republic of China}
\affiliation{Key Laboratory for the Structure and Evolution of Celestial Objects, Chinese Academy of Sciences, 396 Yangfangwang, Guandu District, Kunming, 650216, People’s Republic of China}

\author{En-Wei Liang}
\affiliation{Guangxi Key Laboratory for Relativistic Astrophysics, School of Physical Science and Technology, Guangxi University, Nanning 530004, People's Republic of China}

\author{Alexei V. Filippenko}
\affiliation{Department of Astronomy, University of California, Berkeley, CA 94720-3411, USA}

\begin{abstract}

Mrk 501 is a prototypical high-synchrotron-peaked blazar (HBL) and serves as one of the primary targets for the {\it Imaging X-ray Polarimetry Explorer} ({\it IXPE}). In this study, we report X-ray polarization measurements of Mrk 501 based on six {\it IXPE} observations. The detection of X-ray polarization at a confidence level exceeding 99\% is achieved in four out of the six observations conducted across the entire energy range (2--8 keV) of {\it IXPE}. The maximum polarization degree ($\Pi_{\rm X}$) is measured to be $15.8\%\pm2.8\%$, accompanied by a polarization angle ($\psi_{\rm X}$) of $98.0\degr\pm5.1\degr$ at a confidence level of $5.6 \sigma$. During the remaining two observations, only an upper limit of $\Pi_{\rm X}<$12\% could be derived at the 99\% confidence level. No temporal variability in polarization is observed throughout all six {\it IXPE} observations for Mrk 501. A discernible trend of energy-dependent variation in the polarization degree is detected in optical spectropolarimetry; however, no analogous indication is observed in $\Pi_{\rm X}$. The chromatic behavior of $\Pi$ and the consistent values of $\psi$ across different frequencies from X-rays to radio waves, along with the agreement between $\psi$ and jet position angle, strongly support the interpretation of the energy-stratified model with shock-accelerated particles in the jet of Mrk 501. Additionally, the possibility of the presence of a global helical magnetic field in the jet of Mrk 501 is discussed.

\end{abstract}


\section{Introduction}

Blazars are a subclass of active galactic nuclei (AGNs), including BL Lacertae objects (BL Lacs) and flat-spectrum radio quasars (FSRQs). Their broadband spectral energy distributions (SEDs) generally exhibit a bimodal pattern; the first hump in the infrared--optical--ultraviolet and even X-ray bands is thought to be from the synchrotron emission of relativistic electrons in the jet, while the origin of the second bump in MeV--GeV--TeV bands remains subject to debate, usually attributed to the inverse-Compton (IC) scattering process of relativistic electrons \citep[e.g.,][]{1992ApJ...397L...5M,1996A&AS..120C.503G,2009ApJ...704...38S,2012ApJ...752..157Z,2014ApJ...788..104Z,2015ApJ...807...51Z,2014MNRAS.439.2933Y}. Based on the peak frequency of their synchrotron radiation component, blazars are typically categorized into three groups: low-synchrotron-peaked blazars (LBLs; $\nu_{\rm s}<10^{14}$ Hz), intermediate-synchrotron-peaked blazars (IBLs; $10^{14}<\nu_{\rm s}<10^{15}$ Hz), and high-synchrotron-peaked blazars (HBLs; $\nu_{\rm s}>10^{15}$ Hz) \citep{2010ApJ...710.1271A}. Relativistic jets of blazars serve as ideal sites for particle acceleration, where particles can be accelerated through shocks, stochastic processes, and magnetic reconnection (see \citealp{2020NewAR..8901543M} for a review). Through the measurements of linear polarization, one can determine the degree of order of the magnetic field and its average direction relative to the jet axis, thereby probing both radiation and acceleration mechanisms of the high-energy particles in jets \citep[e.g.,][]{2022Natur.611..677L}.

Previous polarization measurements are limited to the optical band and longer wavelengths. The emergence of a new, superluminal component from the radio core of BL Lacertae is revealed by Very Long Baseline Array (VLBA) observations around the time of a TeV $\gamma$-ray flare, while MOJAVE images demonstrate the polarization changes of the radio core before and after the TeV $\gamma$-ray flare \citep{2013ApJ...762...92A}. The optical polarization changes accompanying $\gamma$-ray flares have been observed in several blazars \citep{2010Natur.463..919A,2010ApJ...715..362J,2013ApJ...773..147J,2015ApJ...813...51C,2017A&A...603A..29A,2018RAA....18...40Z}. The polarization variability patterns observed in radio and optical bands for blazars suggest the presence of an ordered magnetic field, which can be attributed to the compression and amplification by shocks or the existence of a helical magnetic field \citep{2005MNRAS.360..869L,2012AJ....144..105H,2015ApJ...813...51C,2020MNRAS.498..599T,2020NewAR..8901543M}. The launch of the {\it Imaging X-Ray Polarimetry Explorer} \citep[{\it IXPE};][]{2022HEAD...1930101W} enables polarimetry measurements of blazars in the 2--8 keV band. The X-ray polarimetry measurements are crucial, as X-ray emission is associated with higher-energy electrons and probes properties in closer proximity to the acceleration site. 

Mrk 501, one of the nearest GeV--TeV BL Lacs \citep[redshift $z=0.034$;][]{1975ApJ...198..261U}, has been widely studied in multiwavelength observations over the past few decades \citep[e.g.,][]{2004ApJ...600..127G,2015A&A...573A..50A,2023ApJS..266...37A,2024A&A...685A.117M}. During its observation period from April 1997 to June 1999 by BeppoSAX, the peak frequency of the synchrotron emission component shifted from 100 keV back to 0.5 keV, accompanied by a decrease in flux \citep{2001ApJ...554..725T,2013ApJ...767....8Z}. Significant spectral variability was observed in the 0.1--150 keV band between September 1996 and October 2001, with X-ray spectra well described by a log-parabolic law function across all flux states \citep{2004A&A...413..489M}. In addition to variations in the X-ray band, significant flux changes have also been detected in other bands \citep[e.g.,][]{1996ApJ...456L..83Q,2015ApJ...798....2S,2017A&A...603A..31A,2019ApJ...870...93A}. Mrk 501 is a typical HBL and also the first blazar observed by {\it IXPE} \citep{2022Natur.611..677L}.

In this paper, we present {\it IXPE} polarimetry measurements of Mrk 501, along with simultaneous optical polarimetry and spectral and temporal observations in multiple bands. Polarization observations and data analysis are described in Section \ref{sec2}, and the results are reported in Section \ref{sec3}. Section \ref{sec4} provides a discussion and our conclusions. Throughout, $H_0=71$~km~s$^{-1}$~Mpc$^{-1}$, $\Omega_{\rm m}=0.27$, and $\Omega_{\Lambda}=0.73$ are adopted.

\section{Polarization Observations and Data Analysis}\label{sec2}

\subsection{IXPE}\label{sec2.1}

Mrk 501 was observed by {\it IXPE} six times between 2022 March 8 and 2023 April 16 (throughout the paper, UTC dates are used), with the net exposure time ranging from $\sim 87$ ks to $\sim 104$ ks. We proceeded coordinates correction to the publicly available Level-2 event files, which store polarization information in the form of Stokes parameters ($I$, $Q$, and $U$) photon-by-photon, to remove the detector pointing misalignment. Based on the onboard calibration data\footnote{\url{https://bitbucket.org/ixpesw/pi\_corr\_caldb/src/master/}}, we proceeded the energy correction (i.e., PI correction) to all three detector units (DUs) for the first two observations using the $xppicorr$ task in the {\it IXPE} internal software $ixpeobssim$ \citep{2022SoftX..1901194B}, customized for {\it IXPE} data analysis and simulations. This PI correction had been automatically proceeded to the Level-2 event files of those observations performed after June 2022. The source region was defined as a circle centered on the radio position of Mrk 501 with a radius of 60$^{\prime\prime}$, while the background region was identified as an annulus with inner and outer radii of 120$^{\prime\prime}$ and 270$^{\prime\prime}$, respectively. This strategy was applied to all six {\it IXPE} observations of Mrk 501.

We first estimated the polarization using a model-independent method with the software $ixpeobssim$. Photons from both the source and background regions were extracted using the $xpselect$ task. The polarization was calculated using the PCUBE algorithm within the $xpbin$ task. We generated polarization cubes for each of the three DUs to extract information such as normalized Stokes parameters ($\mathcal{Q} \equiv Q/I$ and $\mathcal{U} \equiv U/I$), minimum detectable polarization at 99\% significance (MDP$_{99}$), X-ray polarization degree ($\Pi_{\rm X}$), X-ray polarization angle ($\psi_{\rm X}$), and their corresponding uncertainties in the 2--8 keV band. The estimated polarization parameters from the combined three DUs for each {\it IXPE} observation are given in Table \ref{table_pol}.

We cross-checked the polarization parameters using spectropolarimetric analysis with $Xspec$ \citep{1999ascl.soft10005A} in the HEASoft (v.6.30.1) package. We generated Stokes parameter spectra for both the source and background using the PHA1, PHA1Q, and PHA1U algorithms, which convert the $I$, $Q$, and $U$ Stokes parameters of photons into OGIP-compliant pulse-height analyzer (PHA) files (three Stokes parameter spectra per three DUs). The $I$, $Q$, and $U$ spectra were rebinned using the $ftgrouppha$ task in FTOOL, requiring a minimum of 20 counts per spectral channel for the $I$ spectra. A constant energy bin of 0.2 keV was applied to both the $Q$ and $U$ spectra. The spectropolarimetric fit was performed using $\chi^{2}$ statistics. Firstly, the $I$ spectra were fitted with an absorbed power-law model of the form $TBabs\times po$ within $Xspec$. The power-law function is
\begin{equation}\label{PL} 
\frac{dN}{dE} = N_{0}\left(\frac{E}{E_0}\right)^{-\Gamma_{\rm X}}\, , 
\end{equation}
where $E_{0}=1$ keV is the scale parameter of photon energy, $N_0$ is the power-law normalization, and $\Gamma_{\rm X}$ is the photon spectral index \citep{2004A&A...413..489M}. Only the Galactic photoelectric absorption was considered in the $TBabs$ model, where the neutral hydrogen column density was fixed at the Galactic value along the line of sight toward Mrk 501, $N_{\rm H} = 1.69 \times 10^{20}$ cm$^{-2}$ \citep{2016A&A...594A.116H}. During the fit of the $I$ spectra, both $N_0$ and $\Gamma_{\rm X}$ were allowed to vary. Secondly, we simultaneously fit 3 $\times$ $I$, $Q$, and $U$ spectra using an absorbed power-law model with a constant polarization of the form $TBabs$ $\times$ ($polconst \times po$) within $Xspec$. The polarization model $polconst$, assumes constant polarization parameters within the operating energy range and has only two free parameters, $\Pi_{\rm X}$ and $\psi_{\rm X}$. For the spectropolarimetric fit, we fixed the values of $N_0$ and $\Gamma_{\rm X}$ obtained through fitting the $I$ spectra, with only $\Pi_{\rm X}$ and $\psi_{\rm X}$ being free parameters (see also \citealp{2024ApJ...963....5E}). The results of the spectropolarimetric fits are also presented in Table \ref{table_pol}. If the estimated value of $\Pi_{\rm X}$ was lower than the corresponding MDP$_{99}$, an upper limit at 99\% confidence level would be given using the $error$ task within $Xspec$.

\subsection{Kast Optical Spectropolarimetry}\label{sec2.2}

We followed the methodology outlined by \citet{2022MNRAS.515..138P} for observations and data reduction. We conducted six rounds of spectropolarimetry between 2022 April and 2022 November, using the polarimetry mode of the Kast double spectrograph (Kast) mounted on the Shane 3~m telescope at Lick Observatory\footnote{\url{https://www.lickobservatory.org/}}. 

Each night, we took 600~s exposures at retarder-plate angles of $0\degr$, $45\degr$, $22.5\degr$, and $67.5\degr$ to calculate the Stokes parameters $Q$ and $U$. These observations were done under low-airmass conditions ($\lesssim 1.5$), enabling us to align the slit north–south (at a position angle of $180\degr$) on all nights. Since Kast does not have an atmospheric dispersion compensator, either observing at low airmass or aligning the slit along the parallactic angle is crucial \citep[see][]{1982PASP...94..715F}.

Depending on their observability on a given night, we also observed unpolarized standard stars HD~212311, HD~154892, and HD~57702 on all nights, finding an average Stokes $Q$ and $U$ of $<0.05$\%, confirming low instrumental polarization. In the polarizance test, we found that the Kast spectropolarimeter had a polarimetric response exceeding 99.5\% across the relevant wavelength range 4600--9000~\AA. Furthermore, our observations of two high-polarization standard stars (chosen among HD~183143, HD~245310, HD~204827, HD~154445, HD~155528, and HD~43384) on each night demonstrated that the measured fractional polarization and position angle were consistent within 0.1\% and $3\degr$ (respectively) compared to references, affirming the instrument’s accuracy and stability over time \citep{1992AJ....104.1563S,1996AJ....111..856W}.

The intensity-normalized Stokes parameters $\mathcal{Q}$ and $\mathcal{U}$ were used to calculate the fractional polarization 
\begin{equation}
 \Pi = \sqrt{\mathcal{Q}^{2} + \mathcal{U}^{2}}
\end{equation} 
and the polarization position angle 
\begin{equation}
\psi = \frac{1}{2} \arctan\left(\frac{\mathcal{U}}{\mathcal{Q}}\right)\, .
\end{equation}
It is worth noting that $\Pi_{\rm O}$ tends to favor higher polarization owing to its positive definiteness, especially in the low signal-to-noise ratio (S/N) regime. To counter this bias, we followed the approach outlined by \citet{1997ApJ...476L..27W} to determine the debiased polarization degree $\Pi_{\rm O}$ as 
\begin{equation}
   \Pi_{\rm O}  = \left(\Pi - \frac{\sigma_{\Pi}^{2}}{\Pi}\right) \times h(\Pi - \sigma_{\Pi})
\end{equation}
and polarization angle $\psi_{\rm O}$ as
\begin{equation}
    \psi_{\rm O} = \psi\, ,
\end{equation}
where $\sigma_{\Pi}$ represents the $1\sigma$ uncertainty of $\Pi$ and $h$ is the Heaviside step function. For additional details into the polarization calculation, see \citet{2022MNRAS.515..138P} and \citet{2022MNRAS.509.4058P}.

\section{Results}\label{sec3}

\subsection{Polarization in the 2--8 keV Band}

We conducted a systematic analysis of the observational data obtained from Mrk 501 using {\it IXPE} to investigate the X-ray polarization of the source. Two different methods were employed to estimate the polarization parameters, yielding consistent results within their respective uncertainties, as presented in Table \ref{table_pol}. Of the six observations, four showed a detection of polarization with a confidence level exceeding 99\% across the entire energy band (2--8 keV) covered by {\it IXPE}, indicating that $\Pi_{\rm X}$ is higher than the corresponding MDP$_{99}$ value. However, only the second and sixth observations exhibited the estimated polarization degrees at a confidence level exceeding 5.0$\sigma$. Specifically, for these two observations, we found $\Pi_{\rm X} = 11.4\% \pm 1.8\%$ with $\psi_{\rm X} = 115.3\degr \pm 4.4\degr$ at a confidence level of 6.5$\sigma$ and $\Pi_{\rm X} = 15.8\% \pm 2.8\%$ with $\psi_{\rm X} = 98.0\degr \pm 5.1\degr$ at a confidence level of 5.6$\sigma$, respectively. Moreover, $\Pi_{\rm X}=15.8\%$ represents the highest polarization degree detected by {\it IXPE} in the 2--8 keV band for Mrk 501. The estimated polarization degree for the fourth {\it IXPE} observation ($\Pi_{\rm X} = 9.9\%$) slightly exceeds its MDP$_{99}$ value (9.8\%) with a confidence level of $3.1\sigma$.

\begin{figure}
    \centering
    \includegraphics[angle=0, scale=0.35]{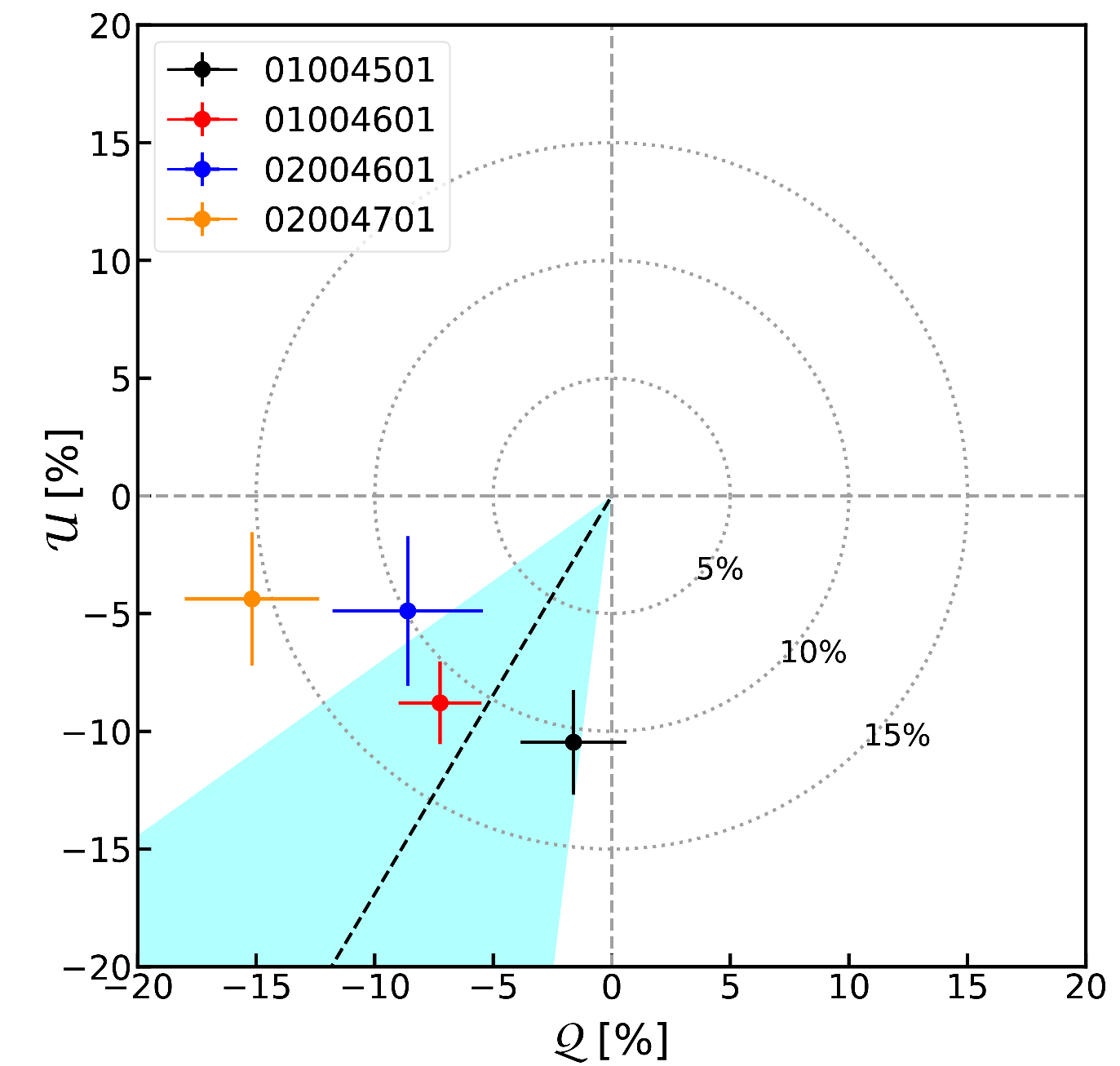}
    \caption{Normalized Stokes parameters $\mathcal{Q}$ and $\mathcal{U}$ in the 2--8 keV band, which were estimated using a model-independent method with the software $ixpeobssim$. The black, red, blue and orange points represent the results of the first, second, fourth and sixth {\it IXPE} observations for Mrk 501, respectively. The black dashed line represents the jet direction of Mrk 501 and the cyan shaded area indicates its associated uncertainty \citep{2022ApJS..260...12W}. The estimated polarization degree of the third and fifth {\it IXPE} observations is lower than its corresponding value of MDP$_{99}$, thus excluded them in the presentation.}
    \label{IXPE}
\end{figure}

In general, the estimated polarization degrees for the four {\it IXPE} observations (with $\Pi_{\rm X}>\rm MDP_{99}$) do not show evident variability given their uncertainties. In addition, the derived values of $\psi_{\rm X}$ demonstrate a decreasing trend from the first to the sixth observations, as displayed in Figure \ref{IXPE} and Table \ref{table_pol}. The values of $\psi_{\rm X}$ obtained during the first, second, and fourth {\it IXPE} observations are consistent with the jet position angle of Mrk 501 within their uncertainties, where the jet position angle of $119.7\degr \pm 11.8\degr$ was determined through VLBA imaging at 43 GHz \citep{2022ApJS..260...12W}. However, the $\psi_{\rm X}$ value is slightly smaller than the jet position angle for the sixth {\it IXPE} observation.  

\subsection{Optical Polarization}
\label{opt-pola}
Using Kast, we performed six optical spectropolarimetric observations of Mrk 501 that were quasi-simultaneous with the second and third {\it IXPE} observations, as described in Section \ref{sec2.2}. The spectrum, normalized Stokes parameters $\mathcal{Q}$ and $\mathcal{U}$, polarization degree $\Pi_{\rm O}$, and polarization angle $\psi_{\rm O}$ are presented in Figure \ref{optical_1} (and Figure \ref{optical_5} in the Appendix), where $\mathcal{Q}$ and $\mathcal{U}$ were rebinned with a bin size of 50~\AA\ to calculate $\Pi_{\rm O}$ and $\psi_{\rm O}$. It is evident that the absolute values of both $\mathcal{Q}$ and $\mathcal{U}$ exhibit a decreasing trend with increasing wavelength. Particularly noteworthy is the clear decrease in the value of $\Pi_{\rm O}$ as wavelength increases; however, no correlation is observed between the variations of $\psi_{\rm O}$ and wavelength. Each of these spectropolarimetric observations consistently demonstrates a reduction in polarization degree with increasing wavelength.

Except for the observation on 2022 July 21, the derived values of $\Pi_{\rm O, ave}$ and $\psi_{\rm O, ave}$ from the other five Kast epochs are approximately consistent, as presented in Table \ref{table_kast} in the Appendix. All $\Pi_{\rm O, ave}$ values are below 5\%, while all $\psi_{\rm O, ave}$ values align with the jet position angle of Mrk 501 \citep{2022ApJS..260...12W} within their uncertainties. The derived values of $\Pi_{\rm O, ave}$ and $\psi_{\rm O, ave}$ obtained from the observation on 2022 July 21 are notably smaller than others. 

The observations of Mrk 501 in the optical band are well known to be significantly influenced by the starlight of its host galaxy \citep{1999ApJ...512...88U,2000ApJ...532..816U,2000ApJ...542..731F,2000ApJ...544..258S}. The starlight is commonly considered to be unpolarized. In order to obtain the intrinsic optical polarization degree $\Pi_{\rm O, intr}$, we had corrected for the depolarization effect caused by the host galaxy using a procedure described in the Appendix. However, due to the lack of contemporaneous optical observations for estimating the total flux density of Mrk 501, we were only able to apply this correction to the estimated polarization degree on 2022 April 1, which was quasi-simultaneous with the second {\it IXPE} observation. Our analysis yielded an intrinsic optical polarization degree of $\Pi_{\rm O, intr} = 5.1\% \pm 1.1\%$ with $\psi_{\rm O} = 109.4\degr \pm 3.0\degr$. Our findings are consistent with the previously reported values of $\Pi_{\rm O} = 5\% \pm 1\%$ and $\psi_{\rm O}=119\degr \pm 9\degr$ as presented by \citet{2022Natur.611..677L}.

\subsection{Energy-Dependent Variability of Polarization}

To investigate whether there are energy-dependent variations in X-ray polarization, we separated the entire 2--8 keV energy range of {\it IXPE} into two smaller ranges (i.e., 2--5 keV and 5--8 keV) and performed spectropolarimetric fits to the Stokes $I$, $Q$, and $U$ spectra, with only $\Pi_{\rm X}$ and $\psi_{\rm X}$ as free parameters (see Section \ref{sec2.1} and \citealp{2024ApJ...963....5E}). As a crosscheck, we also utilized the software $ixpeobssim$ to estimate the polarization parameters in the two smaller ranges for the six {\it IXPE} observations. The derived values of polarization parameters are given in Table \ref{table_pol}, while the $I$, $Q$ and $U$ spectra in 2--5 keV and 5--8 keV for the first {\it IXPE} observation are presented in Figure \ref{IXPE_spec} in the Appendix.

According to the spectropolarimetric fit results, no polarization with $\Pi_{\rm X}>\rm MDP_{99}$ was detected in the 2--8 keV band for the third and fourth {\it IXPE} observations. However, the source exhibited polarization at a $>99\%$ confidence level in the 2--5 keV band during the two observations. No detection of X-ray polarization at a $>99\%$ confidence level  was found across any energy bin for the fifth {\it IXPE} observation.

The polarization parameters can be estimated with a confidence level exceeding $99\%$ in both the 2--5 keV and 5--8 keV bands, but only for the first and second {\it IXPE} observations, as shown in Table \ref{table_pol}. $\Pi_{\rm X}=9.0\%\pm2.0\%$ with $\psi_{\rm X}=138.3\degr\pm6.3\degr$ in the 2--5 keV band and $\Pi_{\rm X}=23.2\%\pm6.1\%$ with $\psi_{\rm X}=126.7\degr\pm7.7\degr$ in the 5--8 keV band were obtained for the first {\it IXPE} observation, while they were $\Pi_{\rm X}=9.2\%\pm1.6\%$ with $\psi_{\rm X}=112.5\degr\pm4.9\degr$ in the 2--5 keV band and $\Pi_{\rm X}=21.1\%\pm4.4\%$ with $\psi_{\rm X}=119.6\degr\pm5.9\degr$ in the 5--8 keV band for the second {\it IXPE} observation. We found that the $\Pi_{\rm X}$ value in the 5--8 keV band overlaps with that in the 2--5 keV band within their respective 2$\sigma$ uncertainties during both the first and second {\it IXPE} observations. No tendency of energy-dependent variation in polarization degree, similar to the optical spectropolarimetric observations, was observed during the first and second {\it IXPE} observations. Hence, no indication of energy-dependent variation in X-ray polarization was observed for Mrk 501, as illustrated in Figure \ref{pol_contour} in the Appendix.

The chromatic behavior of $\Pi$, with $\Pi_{\rm X}>\Pi_{\rm O}>\Pi_{\rm R}$ and consistent polarization angles in three bands, has been reported for Mrk 501 during its first and second {\it IXPE} observations \citep{2022Natur.611..677L} and other HBLs \citep{2022ApJ...938L...7D,2023ApJ...953L..28M,2024arXiv240601693K}. As there are no other simultaneous optical polarization observations available for subsequent {\it IXPE} observations, a comparison between the values of $\Pi_{\rm X}$ and $\Pi_{\rm O}$ cannot be made. However, the quasi-simultaneous radio observations at 43 GHz were conducted during the third, fourth, and sixth {\it IXPE} observations as part of the VLBA-BU Blazar Monitoring Program\footnote{\url{http://www.bu.edu/blazars/VLBAproject.html}}. The program details can be found in \citet{2016Galax...4...47J}, \citet{2017ApJ...846...98J} and \citet{2022ApJS..260...12W}. The ratio of $P_{\rm peak}$ to $I_{\rm peak}$ calculated using the values provided on the website can serve as a representative indicator of polarization degree ($\Pi_{\rm R}$) for the radio core (see \citealp{2024A&A...681A..12K,2024ApJ...963....5E}), where $I_{\rm peak}$ and $P_{\rm peak}$ are the peaks of the total and polarized intensity \citep{2005AJ....130.1418J}. Comparing with the X-ray polarization parameters, $\Pi_{\rm R}$ is lower than $\Pi_{\rm X}$ while the polarization angles remain consistent in both bands, as displayed in Figure \ref{lightcurve}. The chromatic behavior of $\Pi$ with similar $\psi$ across different frequencies depicted in Figure \ref{lightcurve}, along with consistent values between $\psi$ and jet position angle, impeccably supports the interpretation of the energy-stratified model with shock-accelerated particles \citep{2022Natur.611..677L,2022ApJ...938L...7D,2023ApJ...953L..28M}. The recent study conducted by \citet{2024A&A...685A.117M} has presented simultaneous polarization measurements at both radio and optical bands during the third {\it IXPE} pointing towards Mrk 501. Their findings also support the energy-stratified model in Mrk 501.

Note that the optical spectropolarimetric observation on July 21, 2022 yielded a polarization angle of $\psi_{\rm O}\sim68.1\degr\pm11.3\degr$, which significantly differs from the quasi-simultaneously observed values at 43 GHz: $\psi_{\rm R}\sim121\degr$ on 2022 July 15 and $\psi_{\rm R}\sim116\degr$ on 2022 July 22 (taken from the VLBA-BU Blazar Monitoring Program). Recently, \cite{2024A&A...685A.117M} also indicated a notable discrepancy in polarization angle between radio and optical bands. However, further multiwavelength simultaneous observations are still required to thoroughly investigate this matter.

\subsection{Temporal Variability of Polarization}

We also investigated the temporal variation of polarization during the six {\it IXPE} observations for Mrk 501. We assessed the time dependence of polarization by determining the null-hypothesis probability with the $\chi^2$ test following the method in \citet{2024A&A...681A..12K}. The PCUBE analysis results of polarization were used. The null hypothesis assumes a constant polarization during the {\it IXPE}  observations.

\begin{figure}
    \centering
    \includegraphics[angle=0, scale=0.3]{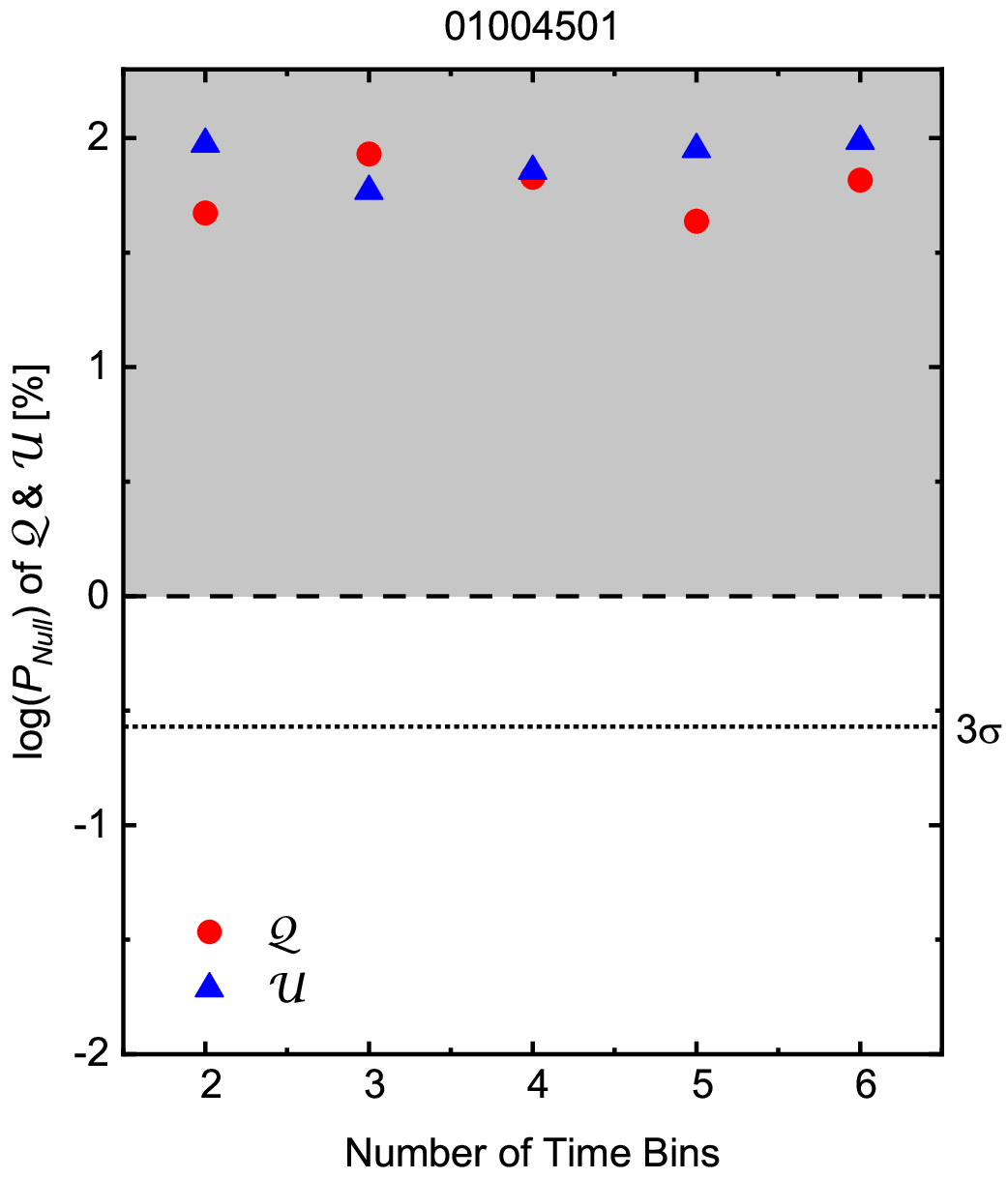}
    \includegraphics[angle=0, scale=0.3]{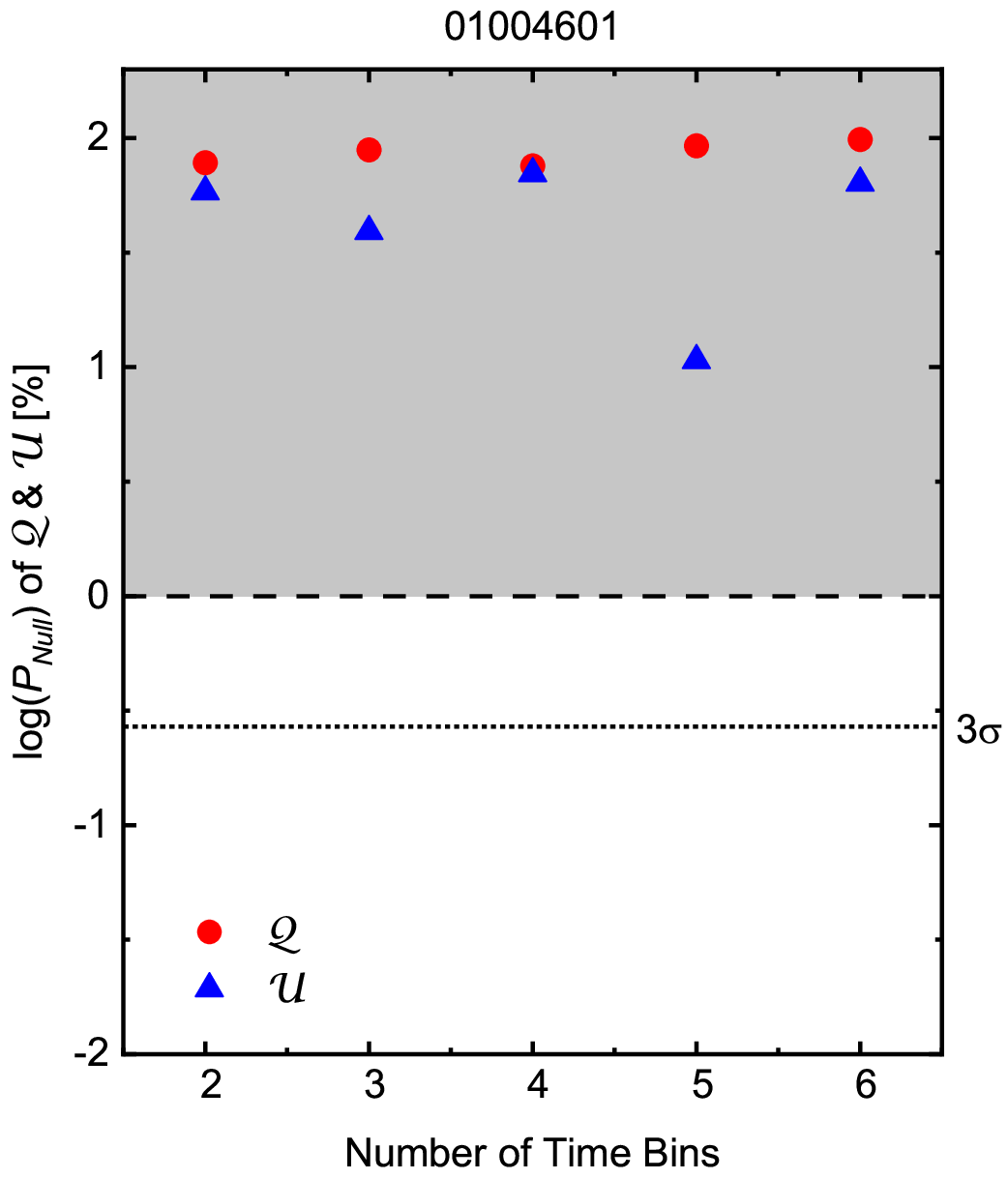}
    \includegraphics[angle=0, scale=0.3]{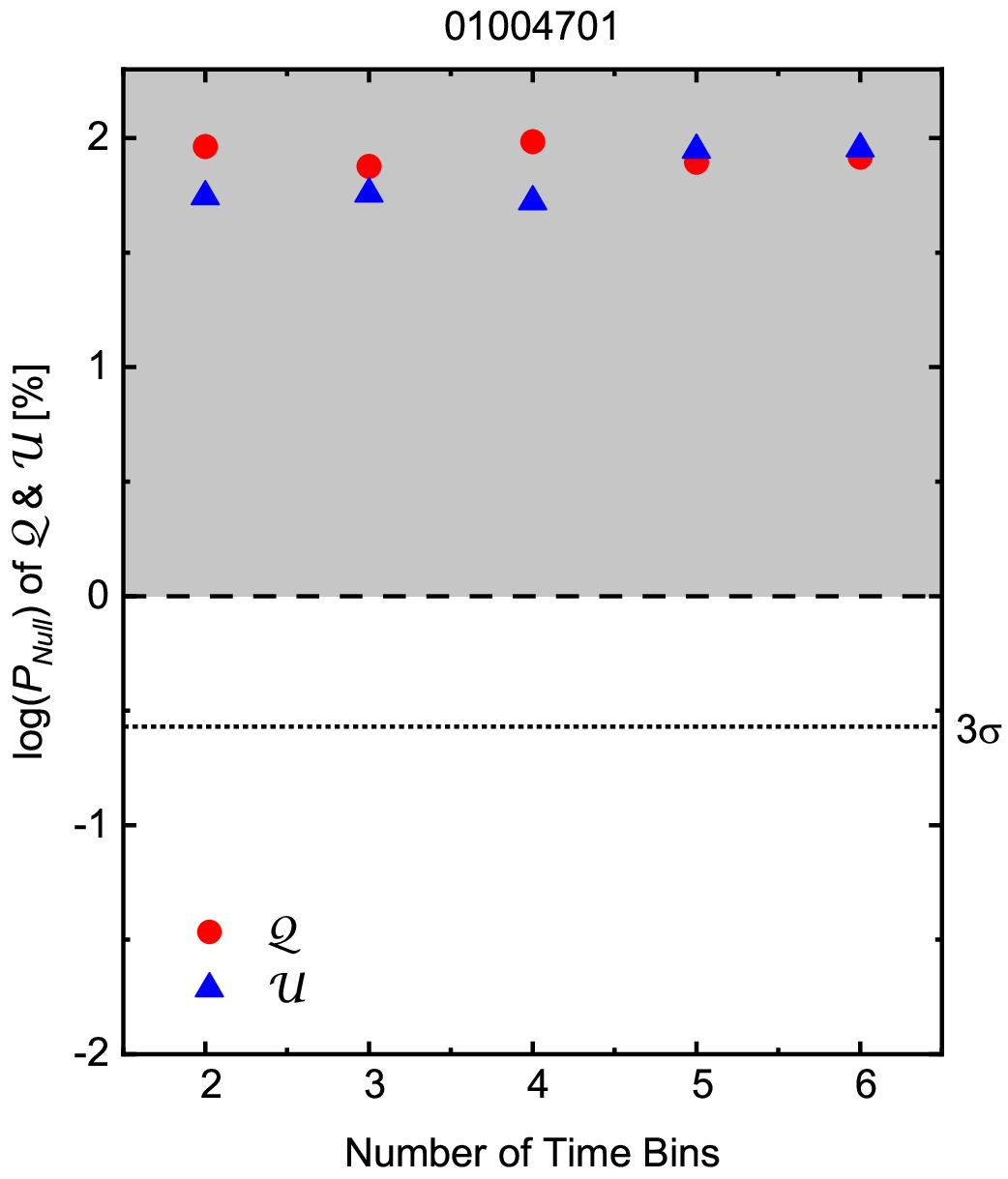}
    \includegraphics[angle=0, scale=0.3]{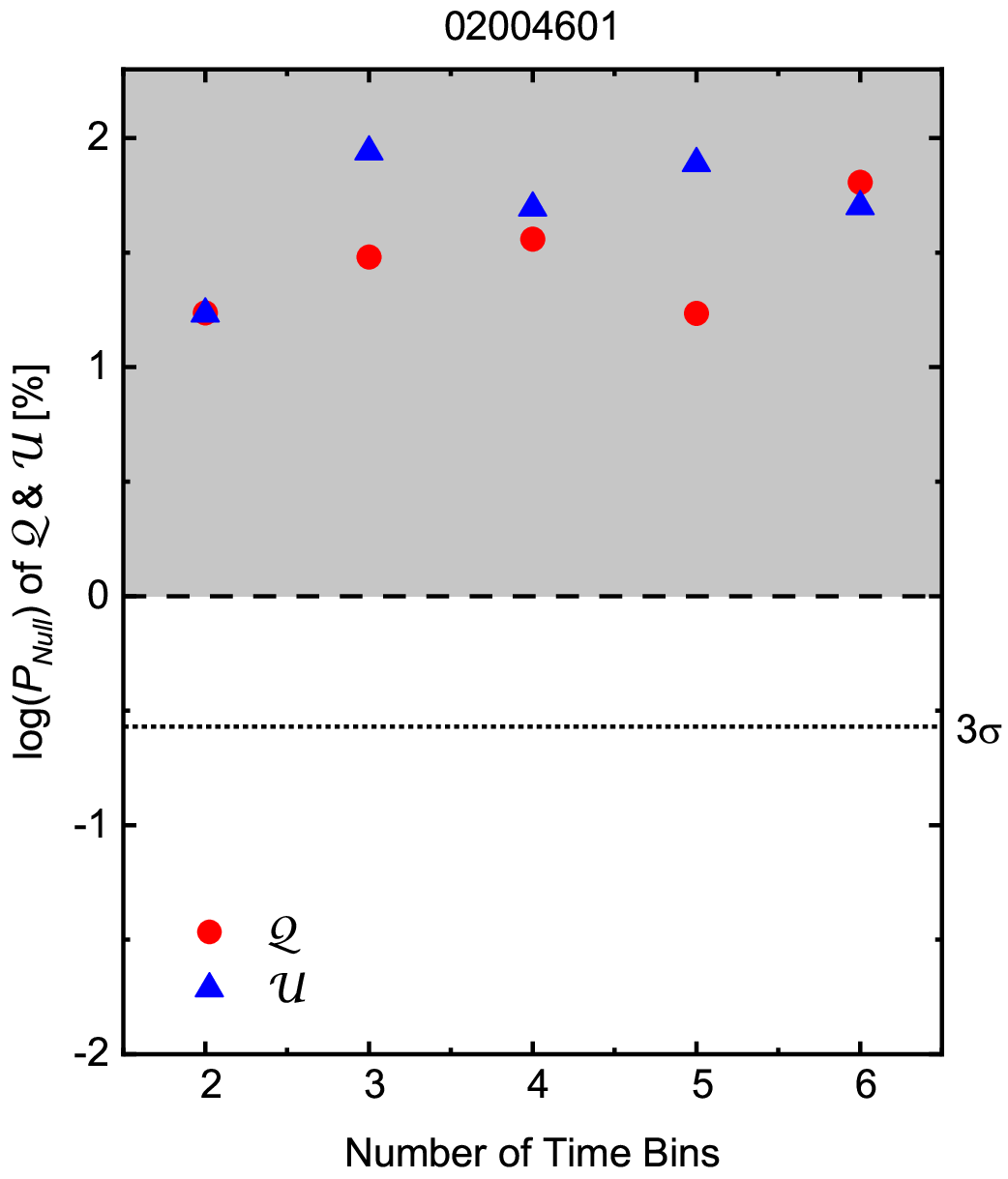}
    \includegraphics[angle=0, scale=0.3]{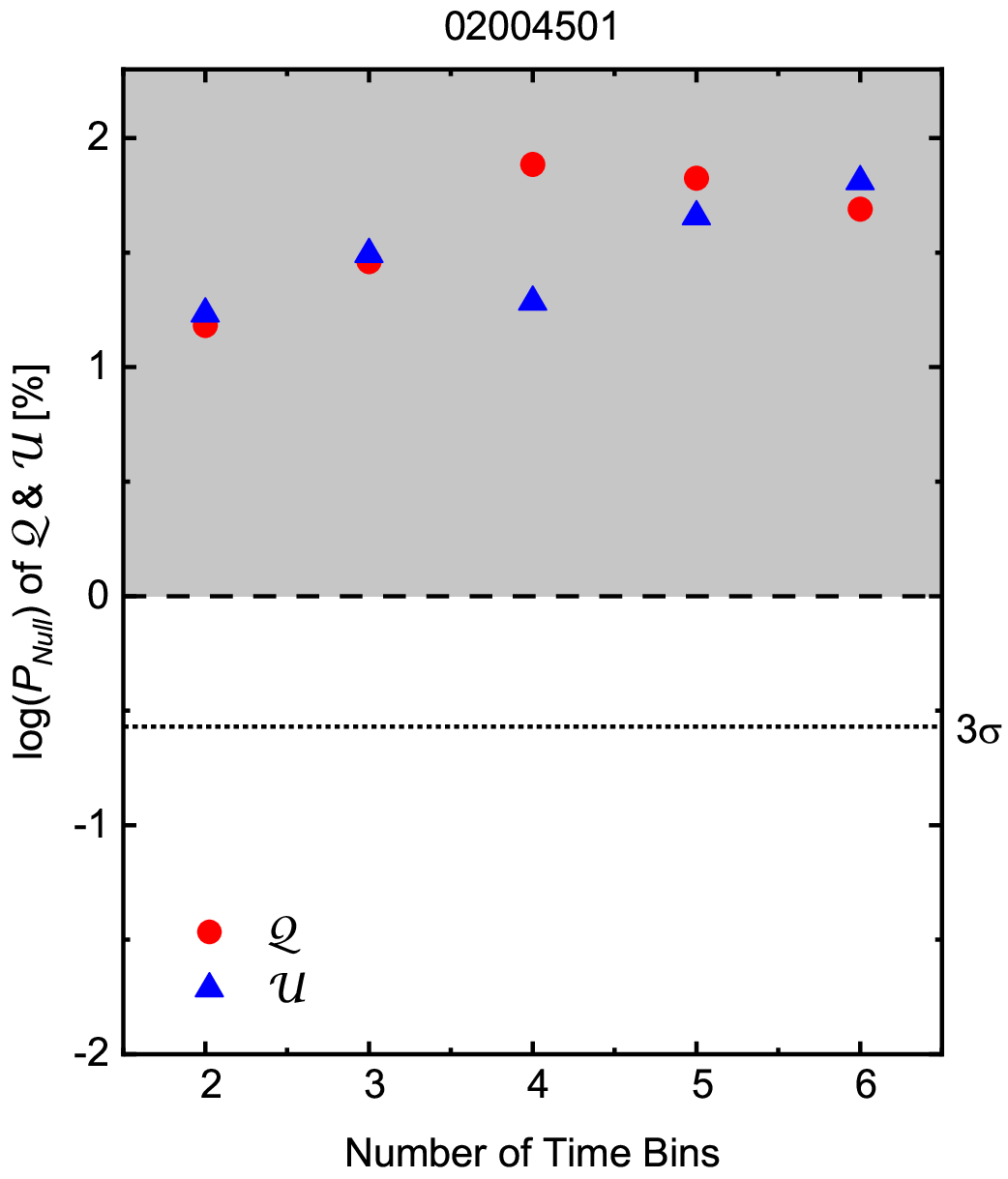}
    \includegraphics[angle=0, scale=0.3]{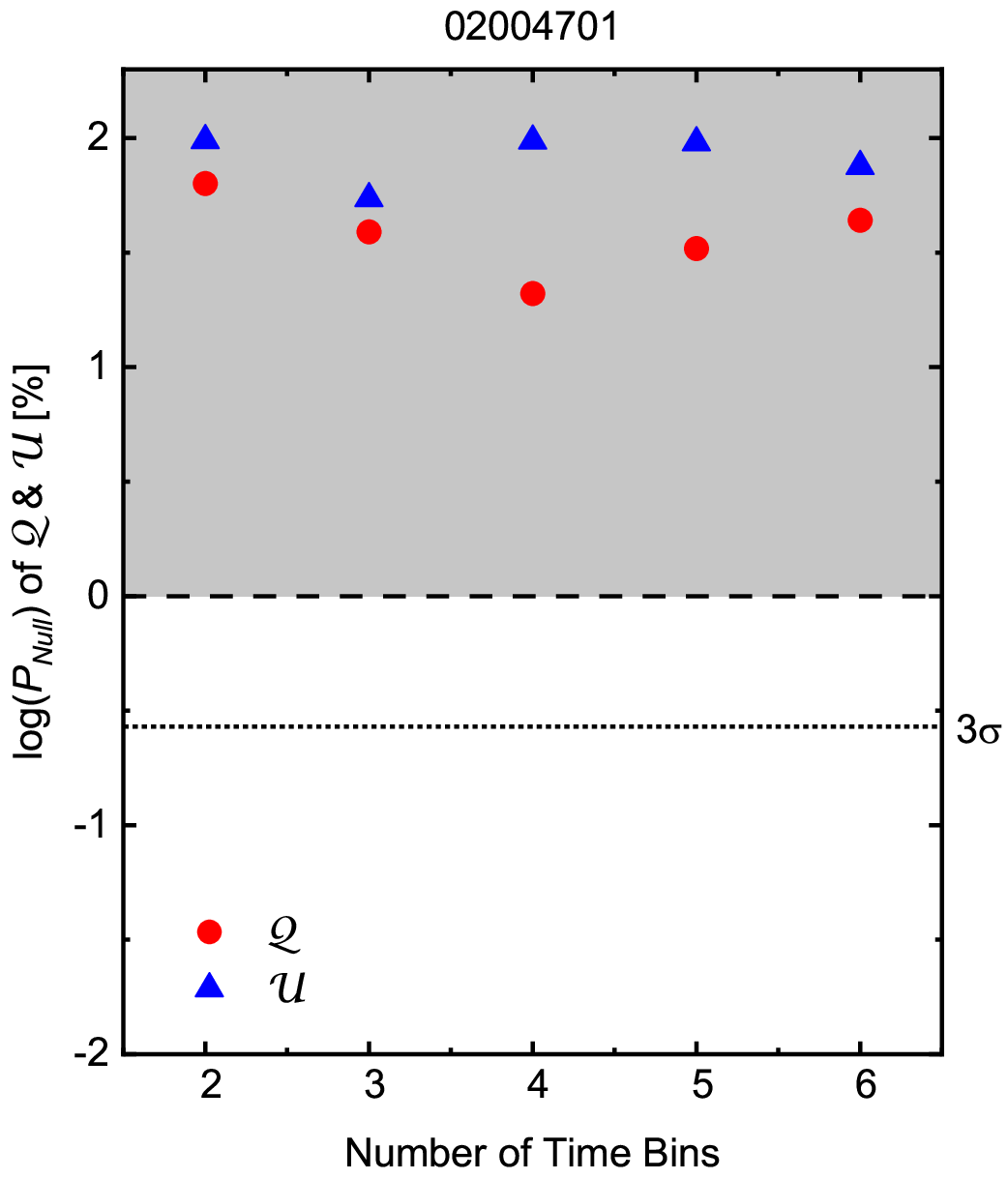}
    \caption{Null-hypothesis probability as a function of the number of time bins for the six {\it IXPE} observations of Mrk 501, followng the method in \cite{2024A&A...681A..12K}. The X-axis represents the number of time bins. The Y-axis represents the null-hypothesis probability in logarithmic scale of the $\chi^{2}$ test for time-dependent variation of the normalized Stokes parameters $\mathcal{Q}$ (red) and $\mathcal{U}$ (blue) with the constant model. The shaded area in gray indicates that the null-hypothesis probability is above the 1\% significance level. The dashed and dotted black lines represent the 1\% and 3$\sigma$ (99.73\%) probability, respectively.}
    \label{stokes_QU}
\end{figure}

Firstly, we divided the one-time observation data into identical time spans that depend on the selected number of bins, and then we measured the normalized Stokes parameter value for each time bin. $\mathcal{Q}$ and $\mathcal{U}$ followed a Gaussian error distribution. Secondly, we compared $\mathcal{Q}$ and $\mathcal{U}$ with the results from fitting each parameter treated as constant during the observation time, i.e., $\mathcal{Q} = \mathcal{Q}_{0}$ and $\mathcal{U} = \mathcal{U}_{0}$. Finally, we calculated the $\chi^{2}$ and the null-hypothesis probability for this case. Each {\it IXPE} observation time was split with 2, 3, 4, 5, 6 time bins, respectively. We depicted the probability of the null hypothesis as a function of the number of time bins for the six {\it IXPE} observations of Mrk 501 in Figure \ref{stokes_QU}. As suggested in \citet{2024A&A...681A..12K}, $P_{\rm Null} >$ 1\% indicates that no temporal variation for $\mathcal{Q}$ or $\mathcal{U}$ was observed statistically during the {\it IXPE} observation, while $P_{\rm Null} <$ 1\% indicates the temporal variation of polarization. As shown in Figure \ref{stokes_QU}, for all six  {\it IXPE} observations, the split $\mathcal{Q}(t)$ and $\mathcal{U}(t)$ light curves with different time bins consistently yield a satisfactory fit with the constant model, i.e., $P_{\rm Null} >$ 1\%. Therefore, no temporal variability of polarization was observed for Mrk 501 during its six {\it IXPE} observations.

\subsection{Flux-Dependent Variability of Polarization}

\begin{figure}
    \centering
    \includegraphics[angle=0, scale=0.5]{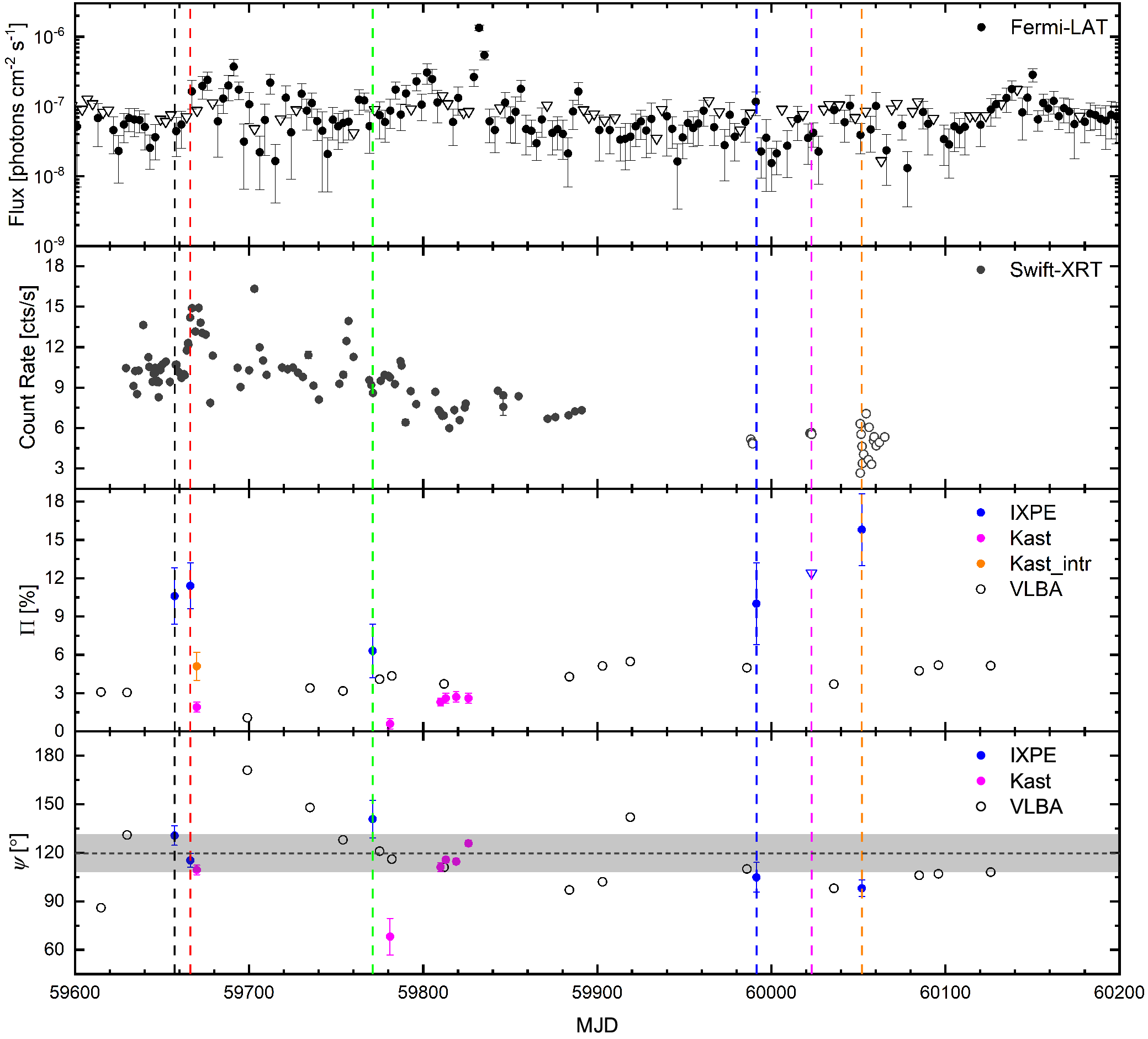}
    \caption{Long-term light curves of Mrk 501 derived from observations with {\it Fermi}-LAT, {\it Swift}-XRT, {\it IXPE}, Kast, and the VLBA. The $\gamma$-ray light curve obtained with {\it Fermi}-LAT observations was binned in intervals of 3 days, and the triangle represents the upper limit value when TS$<$9 for that time bin. The data for the {\it Swift}-XRT count rate were taken from the long-term {\it Swift} monitoring program of {\it Fermi} $\gamma$-ray sources \citep{2013ApJS..207...28S}, where the open circles were derived through our data analysis. In the two bottom panels depicting the $\Pi$ and $\psi$ curves, blue points represent the {\it IXPE} observation results in the 2--8 keV band (for the third {\it IXPE} observation, the $\Pi_{\rm X}$ and $\psi_{\rm X}$ values are that in the 2--5 keV band); magenta points and an orange point (derived intrinsic value) indicate the Kast optical spectropolarimetry; open black circles represent the VLBA observations at 43 GHz taken from the VLBA-BU Blazar Monitoring Program. The horizontal grey dashed line and shaded area in the bottom panel represent the jet direction and the associated uncertainty of Mrk 501, taken from \citet{2022ApJS..260...12W}.}
    \label{lightcurve}
\end{figure}

It is well known that Mrk 501 is a typical HBL and emission in the 2--8 keV band stems from the synchrotron radiation. Therefore, $\Pi_{\rm X}$ serves as an indicator of magnetic-field ordering of the X-ray emission region. The second {\it IXPE} pointing captured a relatively high-flux state of the source in X-rays, as illustrated in Figure \ref{lightcurve}. In contrast, the first and third {\it IXPE} observations revealed moderate-flux states. Notably, the last three {\it IXPE} observations coincided with a period of historically low X-ray flux for Mrk 501. However, due to insufficient statistical data, we are unable to draw a definitive conclusion regarding the correlation between polarization degree and flux in X-rays. We also examined the $\gamma$-ray emission of Mrk 501 using {\it Fermi}-LAT observations during the {\it IXPE} pointings. No flux correlation between the GeV and X-ray bands was observed, which aligns with the expectations of the one-zone leptonic radiation model for HBLs. The emission in the two bands originates from distinct radiation mechanisms of electrons at different energy levels: high-energy electrons contribute to X-rays through the synchrotron process while low-energy electrons account for $\gamma$-rays via the inverse-Compton process \citep{1998ApJ...509..608T}.

The absence of any variation in $\Pi_{\rm X}$ with increasing X-ray flux suggests the injection of particles into a region characterized by stable magnetic-field properties \citep{2022ApJ...938L...7D}, or alternately, changes in the magnetization and/or size of the emission region \citep{2024A&A...685A.117M}. During the third and fourth {\it IXPE} pointings, the estimated values of $\Pi_{\rm X}$ were smaller than their corresponding MDP$_{99}$ in the entire energy band of {\it IXPE} based on spectropolarimetric fits. However, a polarization detection with $>$99\% confidence was observed within the 2--5 keV energy range. These results possibly suggest the presence of a global helical magnetic-field structure in Mrk 501, which is commonly observed in BL Lacs \citep{2021Galax...9...27M}. It could be attributed to this global magnetic-field structure that significant or suggestive polarization is detected in all {\it IXPE} observations. When plasma crosses a shock, the magnetic field becomes more ordered by compression or amplification in the shock, and thus significant polarization is detected.

\section{Discussion and Conclusions}\label{sec4}

We analyzed six {\it IXPE} observations of Mrk 501 to estimate the X-ray polarization. Out of these, only four observations exhibited $>$99\% confidence level detection of polarization in the entire energy band (2--8 keV) of {\it IXPE}. Among them, the highest polarization degree of $\Pi_{\rm X}=15.8\%\pm2.8\%$ with $\psi_{\rm X}=98.0\degr\pm5.1\degr$ was observed at a confidence level of 5.6$\sigma$ in the sixth {\it IXPE} observation. The six {\it IXPE} observations did not reveal any temporal variation in the X-ray polarization for Mrk 501. A trend of energy-dependent variation was observed in our optical spectropolarimetry obtained using the Kast double spectrograph on the Shane 3~m telescope at Lick Observatory: higher polarization degrees were measured at higher frequencies. Therefore, we conducted an investigation into the energy-dependent variability of X-ray polarization. We divided the entire energy range of {\it IXPE} into two narrower intervals, namely 2--5 keV and 5--8 keV, and estimated the polarization parameters within each restricted energy band for every {\it IXPE} observation. However, no similar behavior of energy-dependent polarization variation was observed in X-rays. Additionally, quasi-simultaneous observations carried out with {\it IXPE}, Kast, and the VLBA at 43 GHz revealed that $\Pi_{\rm X}>\Pi_{\rm O}>\Pi_{\rm R}$, with consistent polarization angles across all three bands (see also \citealp{2022Natur.611..677L,2024A&A...685A.117M}). These findings provide strong support for an energy-stratified model for Mrk 501 \citep{2022Natur.611..677L,2022ApJ...938L...7D,2023ApJ...953L..28M,2024A&A...685A.117M}. 

The XRT spectra of Mrk 501 can be fitted with a power-law function; however, when combining the XRT and {\it NuSTAR} spectra, a log-parabola model was required (see Table \ref{table_xrt} in the Appendix). The presence of a curved X-ray spectrum is commonly observed in synchrotron-dominated HBLs (e.g., \citealp{2004A&A...413..489M,2021MNRAS.507.5690G,2022MNRAS.514.3179M,2023ApJ...953L..28M}). During the first two {\it IXPE} observations of Mrk 501, we obtained $\Gamma_{\rm X}<2$ with the XRT observations and the joint analysis of XRT and {\it NuSTAR} data; however, during the remaining four {\it IXPE} observations, $\Gamma_{\rm X}$ exceeded 2. This suggests that there was a shift in the peak frequency of the synchrotron emission component as previously observed  \citep{2001ApJ...554..725T,2013ApJ...767....8Z}, demonstrating a tendency toward \emph{harder when brighter} behavior within the XRT band (Figure \ref{Gamma_x-flux} in the Appendix, see also \citealp{2024A&A...685A.117M}). Similar behavior was also observed during {\it IXPE} observations of Mrk 421 \citep{2022ApJ...938L...7D} and PKS 2155--304 \citep{2024ApJ...963L..41H}. It is generally interpreted as the injection of high-energy electrons by a shock \citep{1998A&A...333..452K,2022ApJ...938L...7D,2024ApJ...965...58Z}.

Except for the sixth {\it IXPE} observation, the values of $\psi_{\rm X}$ are coincident with the jet position angle within uncertainties, and the polarization angles obtained from quasi-simultaneous observations in radio, optical, and X-ray bands also exhibited agreement. These results support shock acceleration as the likely dominant particle acceleration mechanism in the jet of Mrk 501 during the {\it IXPE} observations (see also \citealp{2022Natur.611..677L}). A hint of polarization angle rotation is observed in the {\it IXPE} data --- a decrease in $\psi_{\rm X}$. Considering long-term VLBA observations at 43 GHz, a smooth rotation of $\psi$ is clearly evident. 

Recently, an orphan optical polarization swing was observed in PG~1553+113 between two {\it IXPE} observation intervals, which could be caused by turbulence \citep{2023ApJ...953L..28M}. Furthermore, the rotation of $\psi$ in X-rays was first detected in another typical HBL, Mrk 421; however, no evident rotation of $\psi$ was found at either radio or optical bands during the {\it IXPE} observation. This rotation was attributed to a helical magnetic structure in the jet of Mrk 421 \citep{2023NatAs...7.1245D,2024A&A...681A..12K}. We speculate that a global helical magnetic field exists in the jet of Mrk 501. As a result of this coherent ordering of the magnetic field, detectable polarization consistently manifested itself in the X-ray emission from Mrk 501 throughout all {\it IXPE} observations, albeit occasionally confined to a narrow energy range.

\begin{acknowledgments}

We thank the anonymous referee for the valuable suggestions. This study makes use of VLBA data from the VLBA-BU Blazar Monitoring Program (BEAM-ME and VLBA-BU-BLAZAR; http://www.bu.edu/blazars/BEAM-ME.html), funded by NASA through the {\it Fermi} Guest Investigator Program. This work is supported by the National Key R\&D Program of China (grants 2023YFE0117200 and 2023YFE0101200), the National Natural Science Foundation of China (grants 12373042, 12022305, U1938201, 11973050, 12373041, 12133003, and 12393813), the Programme of Bagui Scholars Programme (WXG), and the Yunnan Revitalization Talent Support Program (Yunling Scholar Project). 

A.V.F.'s research group at UC Berkeley acknowledges financial assistance from the Christopher R. Redlich Fund, Sunil Nagaraj, Landon Noll, Sandy Otellini (K.C.P. is a Nagaraj-Noll-Otellini Graduate Fellow in Astronomy), Gary and Cynthia Bengier, Clark and Sharon Winslow, Alan Eustace (W.Z. is a Bengier-Winslow-Eustace Specialist in Astronomy), William Draper, Timothy and Melissa Draper, Briggs and Kathleen Wood, Sanford Robertson (T.G.B. is a Draper-Wood-Robertson Specialist in Astronomy), and numerous other donors.

A major upgrade of the Kast spectrograph on the Shane 3~m telescope at Lick Observatory, led by Brad Holden, was made possible through generous gifts from the Heising-Simons Foundation, William and Marina Kast, and the University of California Observatories. Research at Lick Observatory is partially supported by a generous gift from Google.

\end{acknowledgments}

\bibliography{reference}
\bibliographystyle{aasjournal}

\begin{sidewaystable}
    \scriptsize 
    \setlength{\tabcolsep}{3.5pt}
    \centering
    \caption{Analysis Results of {\it IXPE} Observation Data for Mrk 501}
    \begin{center}
        \begin{tabular}{ccccccccccccccccc}
        \hline
        \hline
        Method\footnote{The methods used to estimate the polarization. ``I'' indicates model-independent polarimetry using the software $ixpeobssim$, while ``II'' indicates spectropolarimetric fit using $Xspec$.} & OBSID\footnote{The unique identification number specifying the {\it IXPE} observation: ``A'' for 01004501 with exposure of 104,004~s, ``B'' for 01004601 with exposure of 86,634~s, ``C'' for 01004701 with exposure of 97,239~s, ``D'' for 02004601 with exposure of 94,749~s, ``E'' for 02004501 with exposure of 102,368~s, and ``F'' for 02004701 with exposure of 94,775~s.} & Date\footnote{The start time of the {\it IXPE} observation.} & $\Pi_{\rm X}$ & $\psi_{\rm X}$ & $\Gamma_{\rm X}$ & Norm\footnote{The power-law normalization is in units of $\times 10^{-2}$ photons keV$^{-1}$ cm$^{-2}$ s$^{-1}$.} & MDP$_{99}$ & Sig & $\Pi_{\rm X}$ & $\psi_{\rm X}$ & MDP$_{99}$ & Sig & $\Pi_{\rm X}$ & $\psi_{\rm X}$ & MDP$_{99}$ & Sig \\
        & & & (\%) & ($\degr$) & & & (\%) & ($\sigma$) & (\%) & ($\degr$) & (\%) & ($\sigma$) & (\%) & ($\degr$) & (\%) & ($\sigma$) \\
        \cmidrule(r){4-9}\cmidrule(r){10-13}\cmidrule(r){14-17}
        & & & \multicolumn{6}{c}{2--8 keV} & \multicolumn{4}{c}{2--5 keV} & \multicolumn{4}{c}{5--8 keV} \\
        \hline
        I & A & 2022.03.08 & $10.6\pm2.2$ & $130.6\pm6.0$ & \nodata & \nodata & 6.8 & 4.8 & $9.0\pm1.9$ & $132.6\pm6.2$ & 5.9 & 4.6 & \nodata & \nodata & 23.1 & \nodata \\
        & B & 2022.03.27 & $11.4\pm1.8$ & $115.3\pm4.4$ & \nodata & \nodata & 5.3 & 6.5 & $9.0\pm1.6$ & $115.1\pm4.9$ & 4.7 & 5.8 & $23.6\pm5.3$ & $115.6\pm6.6$ & 16.1 & 4.4 \\
        & C & 2022.07.09 & \nodata & \nodata & \nodata & \nodata & 7.2 & \nodata & $6.3\pm2.1$ & $134.5\pm9.3$ & 6.3 & 3.1 & \nodata & \nodata & 25.5 & \nodata \\
        & D & 2023.02.12 & $9.9\pm3.2$ & $104.8\pm9.2$ & \nodata & \nodata & 9.8 & 3.1 & \nodata & \nodata & 8.3 & \nodata & \nodata & \nodata & 34.9 & \nodata \\
        & E & 2023.03.19 & \nodata & \nodata & \nodata & \nodata & 8.6 & \nodata & \nodata & \nodata & 7.4 & \nodata & \nodata & \nodata & 30.4 & \nodata \\
        & F & 2023.04.16 & $15.8\pm2.8$ & $98.0\pm5.1$ & \nodata & \nodata & 8.7 & 5.6 & $16.0\pm2.5$ & $101.9\pm4.4$ & 7.5 & 6.5 & \nodata & \nodata & 30.1 & \nodata \\
        \hline
        II & A & 2022.03.08 & $10.2\pm1.9$ & $134.9\pm5.3$ & $2.39\pm0.02$ & $6.54\pm0.15$ & 6.8 & 5.4 & $9.0\pm2.0$ & $138.3\pm6.3$ & 5.9 & 4.5 & $23.2\pm6.1$ & $126.7\pm7.7$ & 23.1 & 3.8 \\
        & B & 2022.03.27 & $10.8\pm1.5$ & $113.8\pm3.9$ & $2.07\pm0.01$ & $8.49\pm0.10$ & 5.3 & 7.2 & $9.2\pm1.6$ & $112.5\pm4.9$ & 4.7 & 5.8 & $21.1\pm4.4$ & $119.6\pm5.9$ & 16.1 & 4.8 \\
        & C & 2022.07.09 & $<11.7$ & \nodata & $2.44\pm0.02$ & $6.44\pm0.16$ & 7.2 & \nodata & $7.6\pm2.1$ & $135.1\pm8.0$ & 6.3 & 3.6 & $<24.0$ & \nodata & 25.5 & \nodata \\
        & D & 2023.02.12 & $<16.3$ & \nodata & $2.49\pm0.03$ & $4.00\pm0.13$ & 9.8 & \nodata & $9.9\pm2.8$ & $98.6\pm8.2$ & 8.3 & 3.5 & $<42.7$ & \nodata & 34.9 & \nodata \\
        & E & 2023.03.19 & $<12.4$ & \nodata & $2.46\pm0.03$ & $4.48\pm0.13$ & 8.6 & \nodata & $<12.9$ & \nodata & 7.4 & \nodata & $<29.8$ & \nodata & 30.4 & \nodata \\
        & F & 2023.04.16 & $17.2\pm2.4$ & $100.1\pm4.0$ & $2.43\pm0.03$ & $4.61\pm0.14$ & 8.7 & 7.2 & $17.1\pm2.7$ & $103.6\pm5.0$ & 7.5 & 6.3 & $<41.8$ & \nodata & 30.1 & \nodata \\
        \hline
        \end{tabular}
    \end{center}
    \label{table_pol}
\end{sidewaystable}

\clearpage

\begin{figure}
    \centering
    \includegraphics[angle=0, scale=0.35]{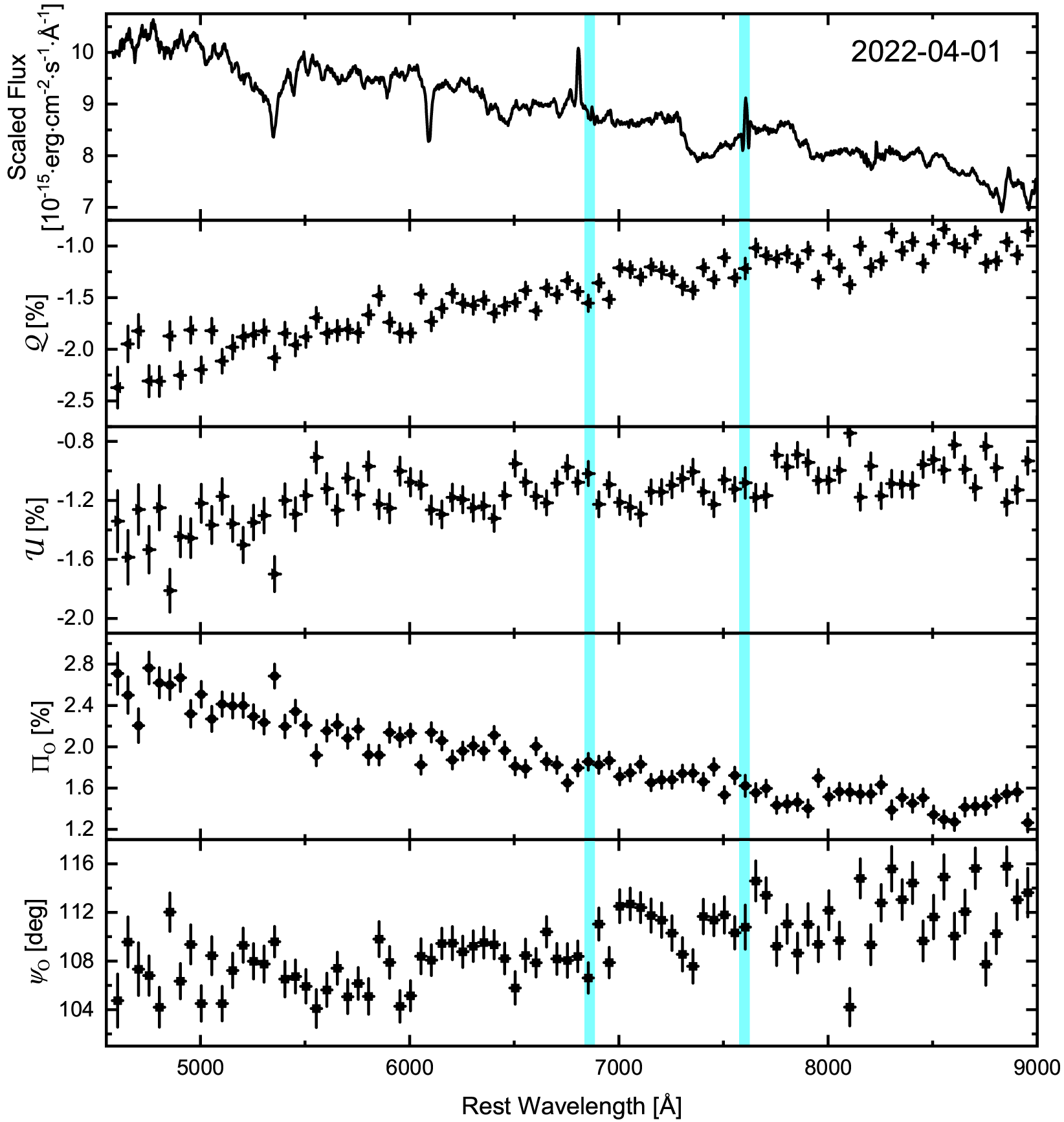}
    \caption{Optical spectropolarimetry of Mrk 501 on 2022 April 1 using the Lick/Kast double spectrograph. From top to bottom: spectrum (line), normalized Stokes parameters $\mathcal{Q}$ (left triangles), $\mathcal{U}$ (right triangles), polarization degree $\Pi_{\rm O}$ (points), and polarization angle $\psi_{\rm O}$ (squares). The two cyan lines located at 6860~\AA\ and 7600~\AA\ represent telluric lines, Earth's atmosphere absorption.}
    \label{optical_1}
\end{figure}

\clearpage

\appendix
\section{Multiwavelength Observations and Data Analysis}

\subsection{Optical}

\subsubsection{Las Cumbres Observatory Global Telescope}

Mrk 501 was observed twice in the $R$-band ($\lambda$ = 6500~\AA, FWHM = 1300~\AA) with exposure of 50~s using the 1~m telescopes with Sinistro cameras in the Las Cumbres Observatory Global Telescope \citep[LCOGT;][]{2013PASP..125.1031B} network during the second {\it IXPE} observation. The raw images were automatically corrected for flat-field and bias effects upon completion of the observations, resulting in the acquisition of two cleaned images. Aperture photometry was carried out following standard routines in the IRAF \footnote{\url{http://ast.noao.edu/data/software/}} package \citep{1986SPIE..627..733T}, using the $apphot$ task. Photometric calibration was performed using the seventh data release of the Sloan Digital Sky Survey \citep[SDSS DR7;][]{2009ApJS..182..543A}, and we converted the $r$-band magnitudes into the $R$-band by applying the transformation equations described in the SDSS DR7 algorithms website \footnote{\url{https://classic.sdss.org/dr7/algorithms/sdssUBVRITransform.php}}. The aperture radius was set to 1.5 $\times$ FWHM (full width at half-maximum intensity) of each individual LCOGT observation. Results of the photometry are presented in Table \ref{table_lcogt}. Note that these data are corrected for Galactic extinction using the Galactic Extinction Calculator\footnote{\url{https://ned.ipac.caltech.edu/forms/calculator.html}}.

\subsubsection{Intrinsic Optical Polarization Degree}

The intrinsic optical polarization degree was estimated by $\Pi_{\rm O, intr} = \Pi_{\rm O, obs} \times I / (I-I_{\rm host})$ \citep{2016A&A...596A..78H}, where $I$ is the total flux density and $I_{\rm host}$ is the flux density of the host galaxy. Since the working wavelength range of Kast is close to the $R$-band, we used the $R$-band data obtained from the LCOGT network to provide an approximate estimation of $I_{\rm host}$ with the photometry aperture radius and FWHM of the source according to the strategy provided by \citet{2007A&A...475..199N}. The values of FWHM for both of the LCOGT observations are also listed in Table \ref{table_lcogt}.

The estimated flux density of the host galaxy is $I_{\rm host} = 8.4 \pm 0.2$ mJy and the average total flux density of Mrk 501 is $I = 13.4 \pm 0.2$ mJy. We therefore obtained $\Pi_{\rm O, intr} = 5.1\% \pm 1.1\%$ with $\psi_{\rm O} = 109.4\degr \pm 3.0\degr$. The effective wavelength range of Kast should be noted as not strictly equivalent to the $R$-band, which may result in a potential inaccuracy in estimating the value of $\Pi_{\rm O, intr}$. Our findings are consistent with the reported values of $\Pi_{\rm O} = 5\% \pm 1\%$ and $\psi_{\rm O} = 119\degr \pm 9\degr$ by \citet{2022Natur.611..677L}. Due to the lack of contemporaneous optical photometry observations for estimating the total flux density $I$ of Mrk 501, we are unable to obtain the intrinsic polarization degrees for the other five spectropolarimetry results by applying this correction.

\subsection{X-rays}

\subsubsection{Swift-XRT} 

The X-ray Telescope \citep[XRT;][]{2005SSRv..120..165B} onboard the {\it Neil Gehrels Swift Observatory} \citep[{\it Swift};][]{2004ApJ...611.1005G} has long been monitoring Mrk 501. During {\it IXPE} observations, XRT performed 21 observations of Mrk 501 in the Windowed Timing (WT) readout mode, which are considered in this work. The data were processed using the XRT Data Analysis Software (XRTDAS, v.3.7.0), developed by the ASI Space Science Data Center (SSDC) and released by the NASA High Energy Astrophysics Science Archive Research Center (HEASARC) in the HEASoft package. The calibration files from XRT CALDB (version 20220803) were used within the $xrtpipeline$ script to calibrate and clean the event files. Events for the spectral analysis were selected from a circle centered on the brightest pixel with a radius of 20 pixels ($\sim 47^{\prime\prime}$), while the background was estimated using an annulus with inner and outer radii of 30 pixels ($\sim 71^{\prime\prime}$) and 45 pixels ($\sim 106^{\prime\prime}$). The ancillary response files (ARFs), which were used to correct the point-spread function (PSF) losses and CCD defects, were generated using the $xrtmkarf$ task with the cumulative exposure maps. The spectra were grouped to ensure a minimum of 20 counts per bin, and then fitted in $Xspec$ with an absorbed single power-law model, as shown in Equation \ref{PL}. In this analysis, we only considered Galactic absorption and fix the neutral hydrogen column density at its Galactic value, similar to what was done in {\it IXPE} spectropolarimetric analysis. We adopted the $\chi^{2}$ minimization technique for all spectral analysis. The best-fit parameters are presented in Table \ref{table_xrt}.

To investigate the X-ray flux state of Mrk 501 during {\it IXPE} observations, we obtained the X-ray light curve taken from a long-term {\it Swift} monitoring program of {\it Fermi} $\gamma$-ray sources\footnote{\url{https://www.swift.psu.edu/monitoring/}} \citep{2013ApJS..207...28S}. The data considered background subtraction, PSF, and pile-up effect corrections, as depicted in Figure \ref{lightcurve}. Note that the observation of Mrk 501 in this program was conducted until 2022 November 8. Therefore, we estimated the count rate value after 2022 November 8 using our data analysis referring to the dark-gray open circle in Figure \ref{lightcurve}.

\subsubsection{NuSTAR}

{\it NuSTAR} has two multilayer-coated telescopes, FPMA and FPMB \citep{2013ApJ...770..103H}. It provides an observational energy band of 3--79 keV, with a spectral resolution of $\sim 1$ keV. During the first two {\it IXPE} observations, {\it NuSTAR} observed Mrk 501. We processed the raw data using the script $nupipeline$ in {\it NuSTAR} Data Analysis Software (NuSTARDAS, v.2.1.2) to obtain calibrated and cleaned files. Source data were extracted from a circle centered on the brightest pixel with a radius of $50^{\prime\prime}$, while the background was estimated using an annulus with inner and outer radii of $100^{\prime\prime}$ and $225^{\prime\prime}$. The spectra were grouped to ensure at least 20 counts per bin, and the spectral analysis was performed using the $\chi^{2}$ minimization technique. The $nuproducts$ package in NuSTARDAS was used to produce the spectra, and the standard {\it NuSTAR} response matrices and effective area files were employed for spectral fits. The log-parabola model with one absorption component provided better fits to the spectra than the single power-law model. The log-parabola function is
\begin{equation}
\frac{dN}{dE}=N_{0}\left(\frac{E}{E_0}\right)^{-(\Gamma_{\rm X}+\beta{\log}(\frac{E}{E_0}))}\, ,
\end{equation}
where $\beta$ is the curvature parameter \citep{2004A&A...413..489M} and $E_0=1$ keV is the scale parameter of photon energy. The neutral hydrogen column density was fixed at the Galactic value as done in the {\it IXPE} spectropolarimetric analysis. Details of the best-fit parameters are presented in Table \ref{table_xrt}.

\subsection{$\gamma$-ray Data}

Mrk 501 has been reported to be associated with the $\gamma$-ray source 4FGL J1653.8+3945 in the {\it Fermi}-LAT latest source catalog \citep[4FGL-DR4;][]{2022ApJS..260...53A,2023arXiv230712546B}. The latest Pass 8 data, covering a period of 15~yr (from 2008 August 4 to 2023 September 30), were downloaded from the Fermi Science Support Center for our analysis. The events in the energy range of 0.1--300 GeV from a region of interest (ROI) centered on the radio position (R.A. = $16^{h}53^{m}52^{s}.217$, decl. = $39^{\circ}45^{\prime}36^{\prime\prime}.609$; J2000) of Mrk 501 with a radius of 15$\degr$ were considered. Data analysis was performed using the publicly available software $fermitools$\footnote{\url{https://fermi.gsfc.nasa.gov/ssc/data/analysis/software/}} (v.2.2.0). The instrument response function of $\rm P8R3\_SOURCE\_V3$ was used. A zenith angle cut of 90$\degr$ was set to avoid $\gamma$-ray contamination caused by Earth's limb. All $\gamma$-ray sources listed in the 4FGL-DR4 within the ROI were added to the XML model for spectral analysis. The spectral parameters of all sources lying within 6$\degr$ were kept free, while the parameters of those sources lying beyond 6$\degr$ were fixed to their 4FGL-DR4 values. The normalization parameters of the Galactic diffuse component (gll\_iem\_v07.fits) and the isotropic emission (iso\_P8R3\_SOURCE\_V3\_v1.txt) were kept free.

The spectrum of Mrk 501 in the 0.1--300 GeV band is well described by a log-parabola model, yielding a photon spectral index $\Gamma_{\gamma}=1.75\pm0.01$ and a curvature parameter $\beta=0.012\pm0.003$. The $\sim 15$~yr average flux of the {\it Fermi}-LAT observations is $F_{\rm 0.1-300~GeV}=(15.40\pm0.39)\times10^{-11}$ erg cm$^{-2}$ s$^{-1}$ with TS = 42519.6. The long-term $\gamma$-ray light curve was extracted with a time bin of 3 days, as illustrated in Figure \ref{lightcurve}.

\clearpage

\begin{deluxetable}{ccccccc}
    \tabletypesize{\small}
    \tablecolumns{7}
    \tablecaption{Kast Spectropolarimetry Results}
    \tablehead{\colhead{Date} & \colhead{$\Pi_{\rm O, max}$} & \colhead{$\psi_{\rm O, max}$} & \colhead{$\Pi_{\rm O, mim}$} & \colhead{$\psi_{\rm O, min}$} & \colhead{$\Pi_{\rm O, ave}$} & \colhead{$\psi_{\rm O, ave}$} \\
    \colhead{} & \colhead{(\%)} & \colhead{(\degr)} & \colhead{(\%)} & \colhead{(\degr)} & \colhead{(\%)} & \colhead{(\degr)}}
    \startdata
    2022.04.01 & $2.8\pm0.2$ & $115.8\pm1.6$ & $1.3\pm0.1$ & $104.1\pm1.6$ & $1.9\pm0.4$ & $109.4\pm3.0$ \\
    2022.07.21 & $1.6\pm0.2$ & $117.0\pm21.9$ & $0.1\pm0.1$ & $44.8\pm4.6$ & $0.6\pm0.4$ & $68.1\pm11.3$ \\
    2022.08.19 & $2.9\pm0.2$ & $117.8\pm1.7$ & $1.8\pm0.1$ & $95.0\pm4.5$ & $2.3\pm0.3$ & $111.1\pm2.7$ \\
    2022.08.22 & $3.7\pm0.2$ & $122.5\pm1.4$ & $2.0\pm0.1$ & $112.1\pm1.1$ & $2.6\pm0.4$ & $115.7\pm1.8$ \\
    2022.08.28 & $3.8\pm0.2$ & $118.8\pm1.2$ & $1.9\pm0.1$ & $110.0\pm1.0$ & $2.7\pm0.4$ & $114.6\pm1.9$ \\
    2022.09.04 & $3.7\pm0.2$ & $132.2\pm1.2$ & $1.9\pm0.1$ & $122.9\pm0.9$ & $2.6\pm0.4$ & $125.8\pm1.8$ \\
    \enddata
\end{deluxetable}\label{table_kast}

\begin{deluxetable}{ccccccc}
    \tabletypesize{\small}
    \tablecolumns{6}
    \tablecaption{Photometry from LCOGT Images}
    \tablehead{\colhead{Date} & \colhead{Band} & \colhead{Exposure} & \colhead{FWHM} & \colhead{Magnitude} & \colhead{Flux Density} \\
    \colhead{} & \colhead{} & \colhead{(s)} & \colhead{(arcsec)} & \colhead{} & \colhead{(mJy)}}
    \startdata
    2022.03.27 & $R$ & 50 & 2.6 & $13.35\pm0.02$ & $13.7\pm0.3$ \\
    2022.03.28 & $R$ & 50 & 2.5 & $13.40\pm0.02$ & $13.1\pm0.2$ \\
    \enddata
\end{deluxetable}\label{table_lcogt}

\begin{deluxetable}{cccccccc}
    \tabletypesize{\scriptsize}
    \tablecolumns{8}
    \tablecaption{Data Analysis Results of the {\it Swift}-XRT and {\it NuSTAR} Observations}
    \tablehead{\colhead{OBSID} & \colhead{Date} & \colhead{Exposure} & \colhead{Mode/Instrument\tablenotemark{\footnotesize{a}}} & \colhead{$\Gamma_{\rm X}$} & \colhead{$\beta$} & \colhead{Flux\tablenotemark{\footnotesize{b}}} & \colhead{$\chi^{\rm 2}$/dof} \\
    \colhead{} & \colhead{} & \colhead{(s)} & \colhead{} & \colhead{} & \colhead{} & \colhead{(10$^{-10}$ erg cm$^{-2}$ s$^{-1}$)} & \colhead{}}
    \startdata
    00096029008 & 2022.03.08 & 844 & WT & $1.94\pm0.03$ & 0 & $3.46\pm0.07$ & 219/222 \\
    00096029009 & 2022.03.08 & 944 & WT & $1.90\pm0.04$ & 0 & $3.50\pm0.07$ & 256/233 \\
    00011184179 & 2022.03.09 & 994 & WT & $1.99\pm0.03$ & 0 & $3.08\pm0.06$ & 267/228 \\
    00011184180 & 2022.03.09 & 1030 & WT & $1.95\pm0.03$ & 0 & $3.13\pm0.06$ & 232/236 \\
    00011184181 & 2022.03.10 & 1084 & WT & $1.95\pm0.03$ & 0 & $2.84\pm0.06$ & 223/213 \\
    00011184182 & 2022.03.10 & 1005 & WT & $1.92\pm0.03$ & 0 & $3.22\pm0.06$ & 274/232 \\
    00011184188 & 2022.03.28 & 975 & WT & $1.80\pm0.02$ & 0 & $5.09\pm0.08$ & 309/288 \\
    00011184189 & 2022.03.29 & 1010 & WT & $1.80\pm0.02$ & 0 & $5.29\pm0.08$ & 355/301 \\
    00011184223 & 2022.07.10 & 904 & WT & $2.10\pm0.03$ & 0 & $2.87\pm0.06$ & 196/209 \\
    00011184224 & 2022.07.11 & 802 & WT & $2.08\pm0.03$ & 0 & $2.68\pm0.06$ & 213/193 \\
    00096558013 & 2023.02.13 & 929 & WT & $2.22\pm0.04$ & 0 & $1.67\pm0.04$ & 200/158 \\
    00015411021 & 2023.02.14 & 923 & WT & $2.22\pm0.04$ & 0 & $1.76^{+0.04}_{-0.05}$ & 171/154 \\
    00096558014 & 2023.02.14 & 896 & WT & $2.14\pm0.04$ & 0 & $1.91\pm0.05$ & 112/154 \\
    00015411036 & 2023.03.19 & 1093 & WT & $2.13\pm0.03$ & 0 & $1.96\pm0.04$ & 225/185 \\
    00015411038 & 2023.03.20 & 1462 & WT & $2.16\pm0.03$ & 0 & $1.98\pm0.04$ & 249/212 \\
    00015411039 & 2023.03.20 & 1110 & WT & $2.13\pm0.03$ & 0 & $1.92\pm0.04$ & 231/186 \\
    00015411051 & 2023.04.17 & 616 & WT & $2.12\pm0.04$ & 0 & $2.22\pm0.06$ & 168/139 \\
    00015411052 & 2023.04.17 & 775 & WT & $2.06\pm0.06$ & 0 & $1.46\pm0.06$ & 80/81 \\
    00015411053 & 2023.04.17 & 1141 & WT & $2.17\pm0.03$ & 0 & $2.47\pm0.05$ & 242/189 \\
    00015411054 & 2023.04.18 & 825 & WT & $2.10\pm0.04$ & 0 & $2.24\pm0.06$ & 152/139 \\
    00015411055 & 2023.04.18 & 818 & WT & $2.15\pm0.05$ & 0 & $1.71\pm0.06$ & 116/100 \\
    00015411056 & 2023.04.19 & 844 & WT & $2.07\pm0.04$ & 0 & $2.40\pm0.07$ & 152/128 \\
    00015411057 & 2023.04.20 & 774 & WT & $2.07\pm0.03$ & 0 & $2.44\pm0.06$ & 219/179 \\
    00015411058 & 2023.04.21 & 290 & WT & $2.10\pm0.09$ & 0 & $2.06^{+0.11}_{-0.12}$ & 62/44 \\
    00015411059 & 2023.04.22 & 949 & WT & $2.16\pm0.03$ & 0 & $2.08\pm0.05$ & 195/180 \\
    00015411060 & 2023.04.23 & 930 & WT & $2.10\pm0.05$ & 0 & $1.74^{+0.05}_{-0.06}$ & 100/110 \\
    00015411061 & 2023.04.24 & 937 & WT & $2.17\pm0.04$ & 0 & $2.11\pm0.05$ & 167/160 \\
    00015411062 & 2023.04.25 & 924 & WT & $2.22\pm0.03$ & 0 & $1.87\pm0.05$ & 187/163 \\
    00015411063 & 2023.04.26 & 1063 & WT & $2.23\pm0.04$ & 0 & $1.69\pm0.04$ & 218/163 \\
    00015411064 & 2023.04.28 & 1300 & WT & $2.17\pm0.03$ & 0 & $1.91\pm0.04$ & 177/189 \\
    00015411065 & 2023.05.01 & 954 & WT & $2.13\pm0.04$ & 0 & $1.89\pm0.05$ & 159/172 \\ 
    \hline
    60701032002 & 2022.03.09 & 19724 & FPMA & $1.82\pm0.11$ & $0.33\pm0.06$ & $1.86\pm0.02$ & 373/394 \\
    & & 19548 & FPMB & $1.65^{+0.11}_{-0.12}$ & $0.41\pm0.06$ & $1.87\pm0.02$ & 393/381 \\
    & & 19636 & FPMA+FPMB & $1.74\pm0.08$ & $0.37\pm0.04$ & $1.85\pm0.01$ & 778/778 \\
    & & \nodata & XRT+FPMA+FPMB & $1.81\pm0.02$ & $0.32\pm0.01$ & \nodata & 1199/1092 \\
    60702062004 & 2022.03.27 & 20297 & FPMA & $1.78\pm0.08$ & $0.25\pm0.04$ & $3.14\pm0.02$ & 487/481 \\
    & & 20143 & FPMB & $1.78\pm0.08$ & $0.26\pm0.04$ & $3.21\pm0.02$ & 435/473 \\
    & & 20220 & FPMA+FPMB & $1.78\pm0.06$ & $0.26\pm0.03$ & $3.17\pm0.01$ & 944/957 \\
    & & \nodata & XRT+FPMA+FPMB & $1.81\pm0.02$ & $0.24\pm0.01$ & \nodata & 1268/1247 \\
    \enddata
    \tablenotetext{a}{WT represents the readout mode of {\it Swift}-XRT, while FPMA and FPMB represent the two multilayer-coated telescopes of {\it NuSTAR}.}
    \tablenotetext{b}{Flux is calculated in the 0.3--10 keV band for {\it Swift}-XRT data and in the 3--79 keV band for {\it NuSTAR} data.}
    \label{table_xrt}
\end{deluxetable}

\clearpage

\begin{figure}
    \centering
    \includegraphics[angle=0, scale=0.225]{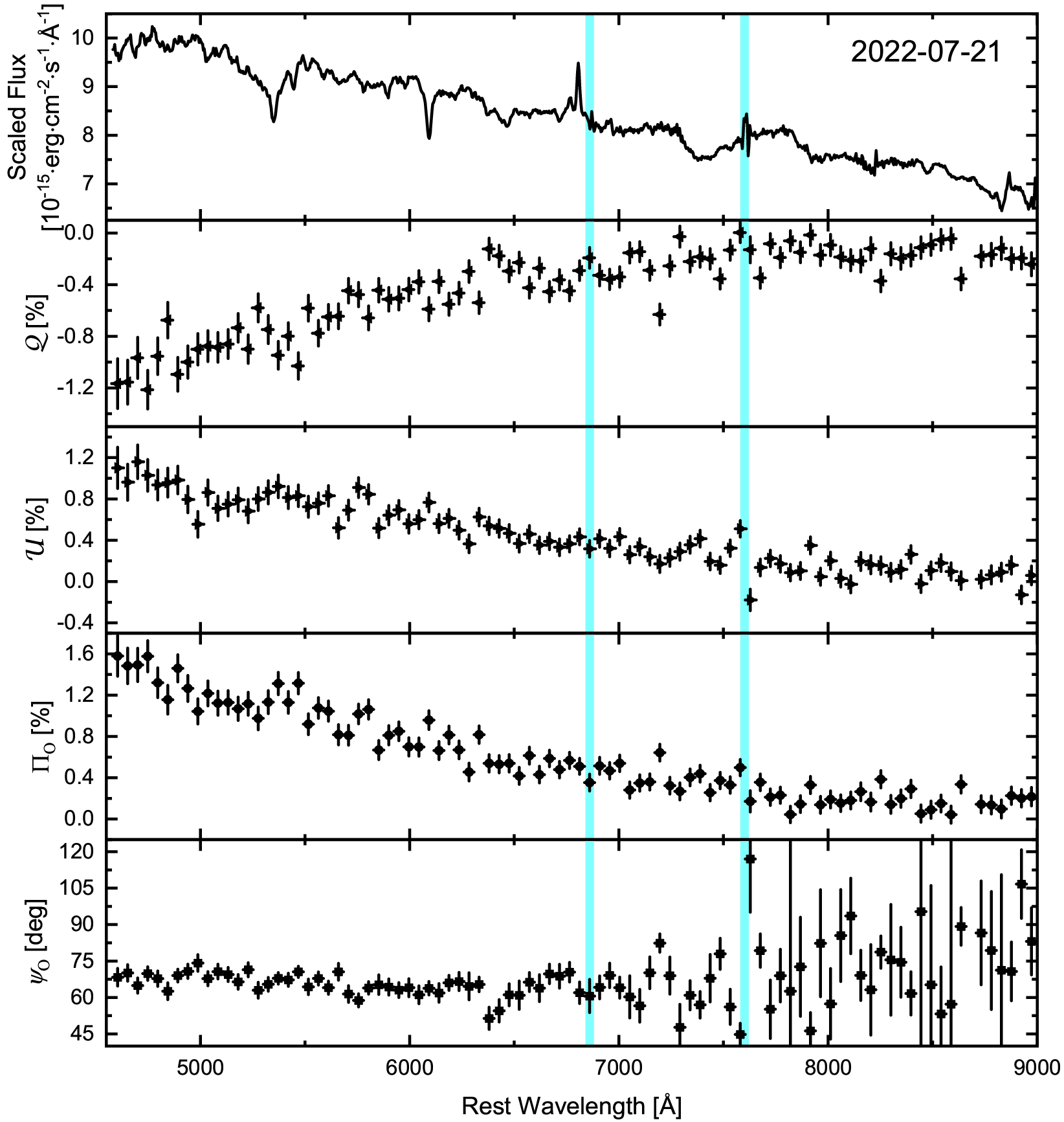}
    \includegraphics[angle=0, scale=0.225]{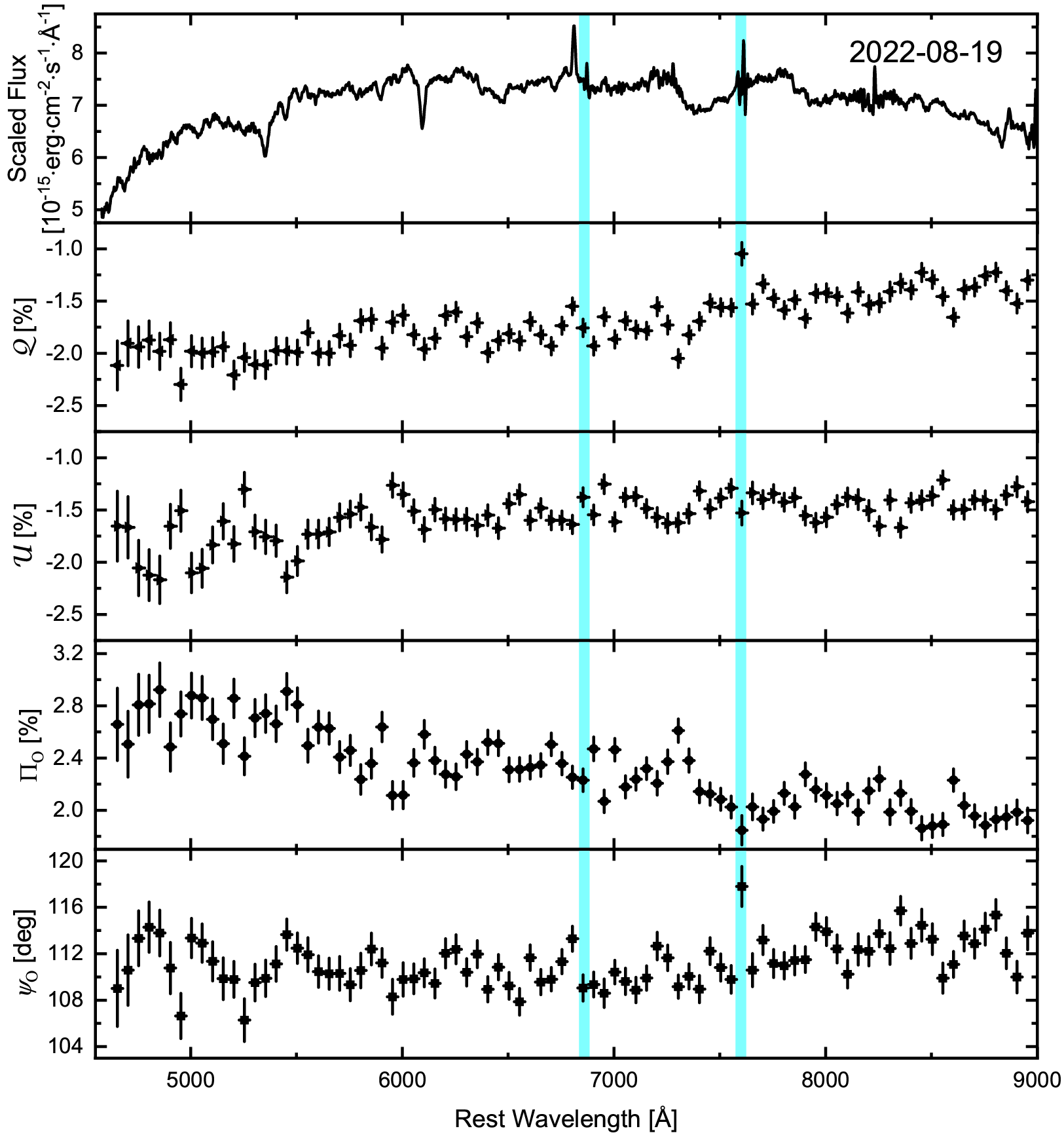}
    \includegraphics[angle=0, scale=0.225]{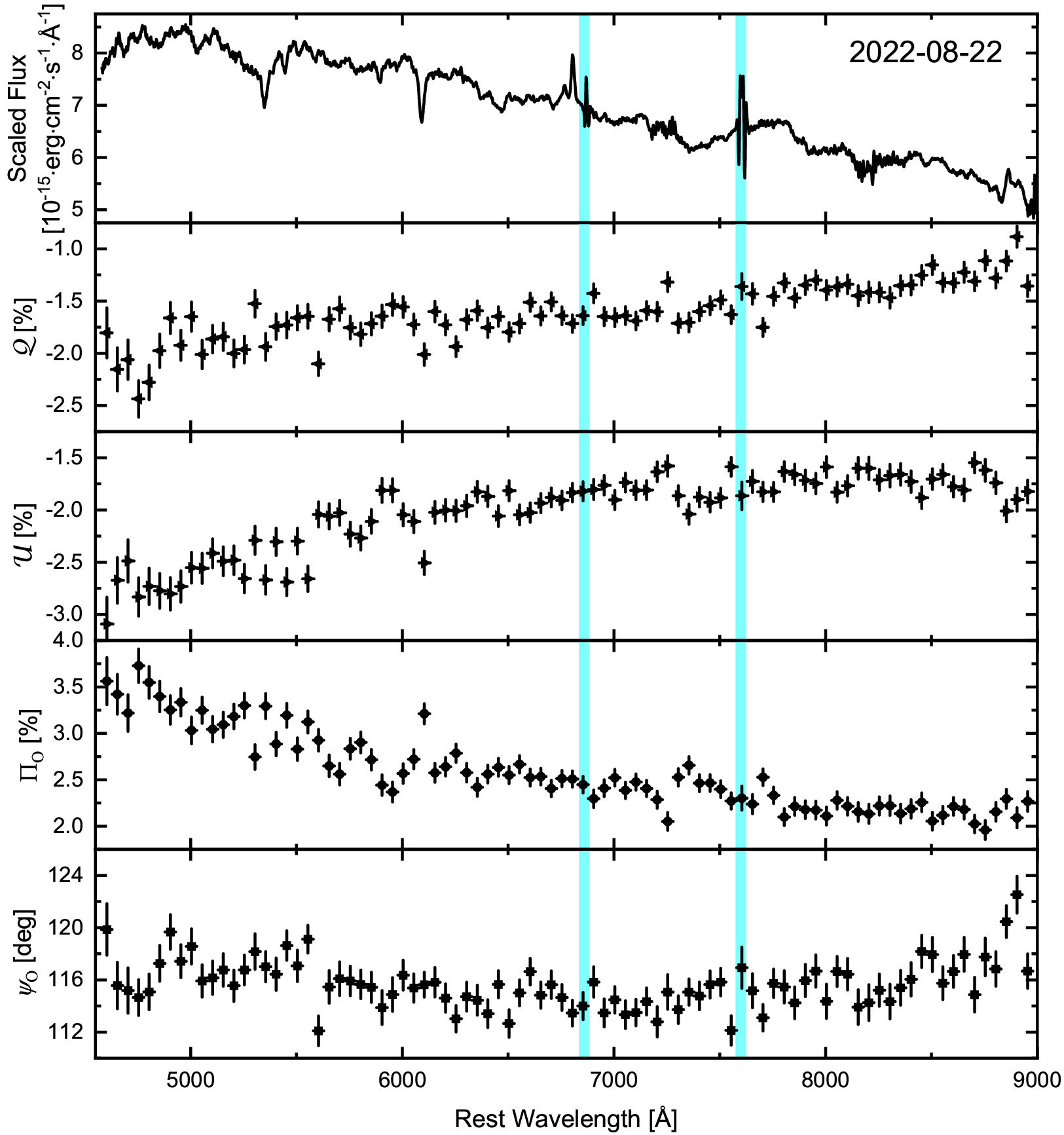}
    \includegraphics[angle=0, scale=0.225]{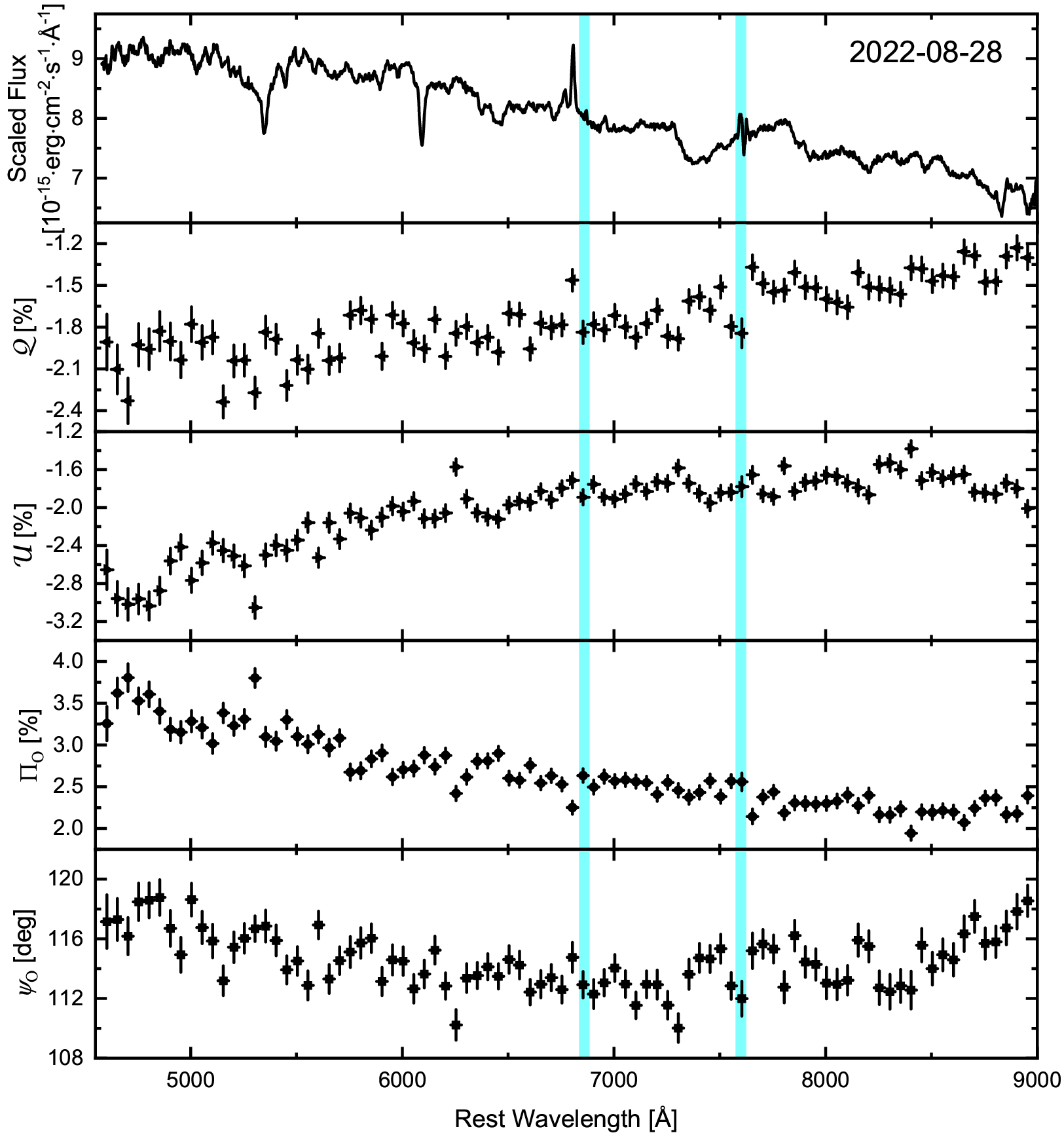}
    \includegraphics[angle=0, scale=0.225]{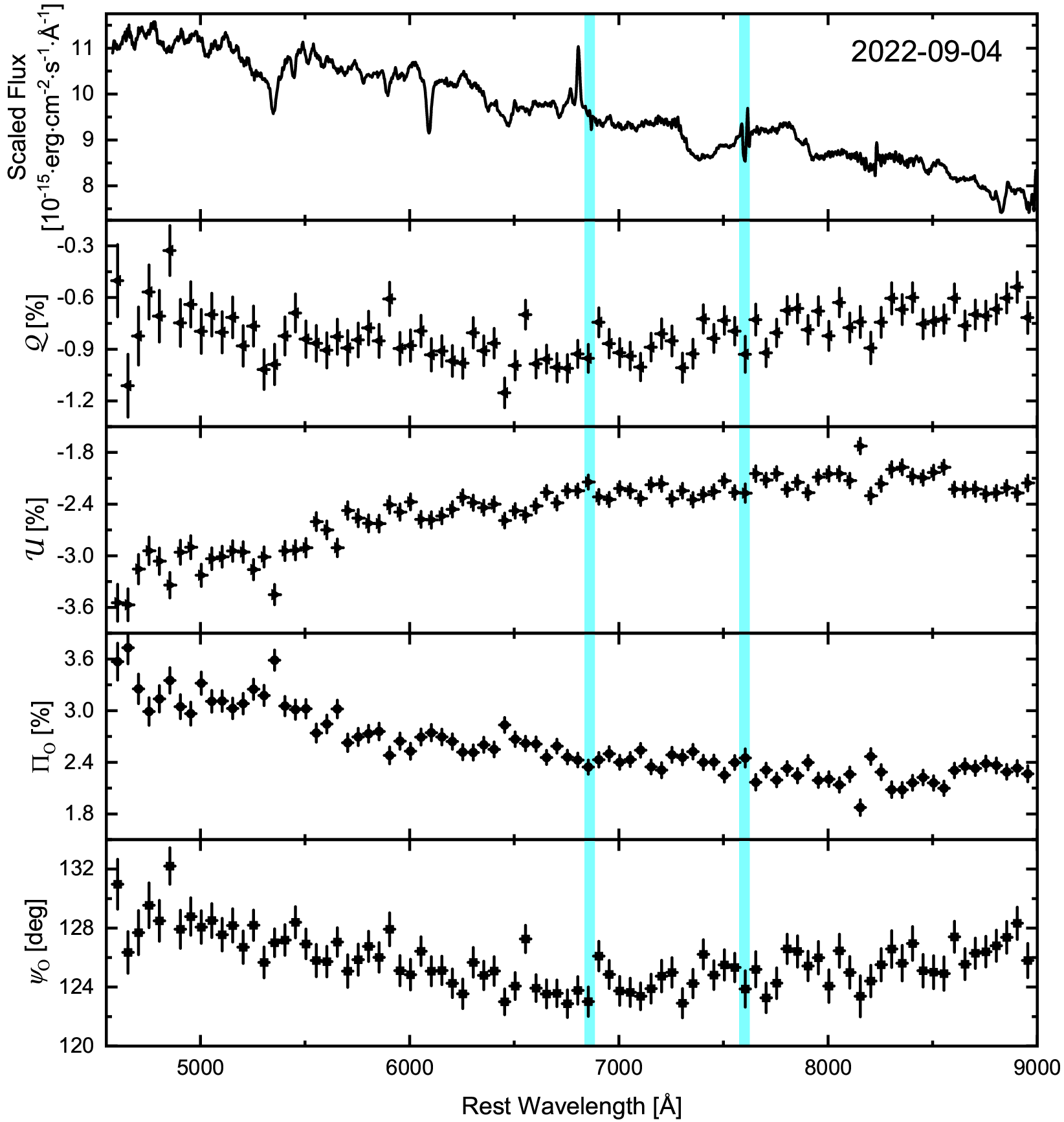}
    \caption{Optical spectropolarimetry of Mrk 501 using the Kast double spectrograph. Symbols are the same as in Figure \ref{optical_1}.}
    \label{optical_5}
\end{figure}

\begin{figure}
    \centering
    \includegraphics[angle=0, scale=0.55]{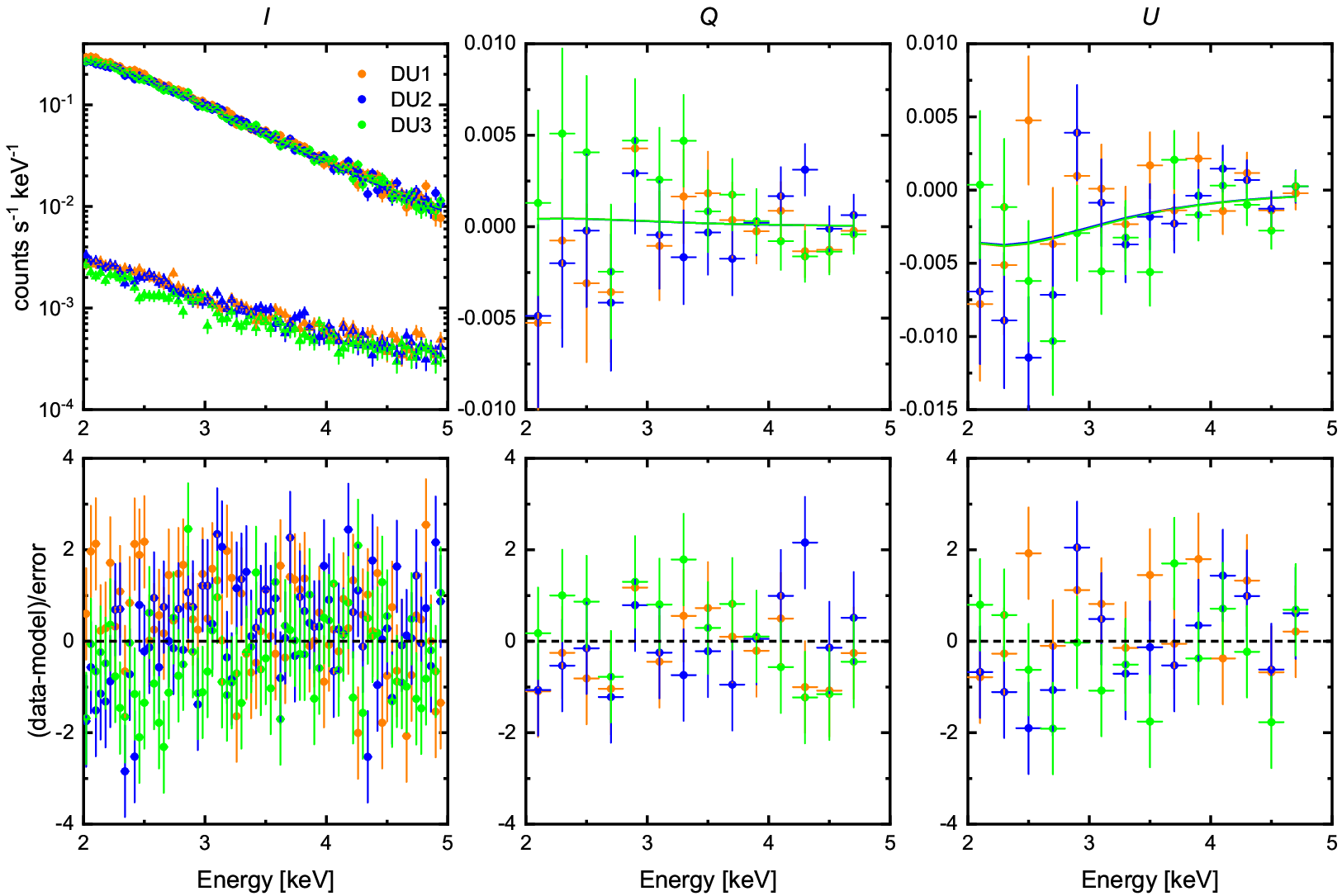}
    \includegraphics[angle=0, scale=0.55]{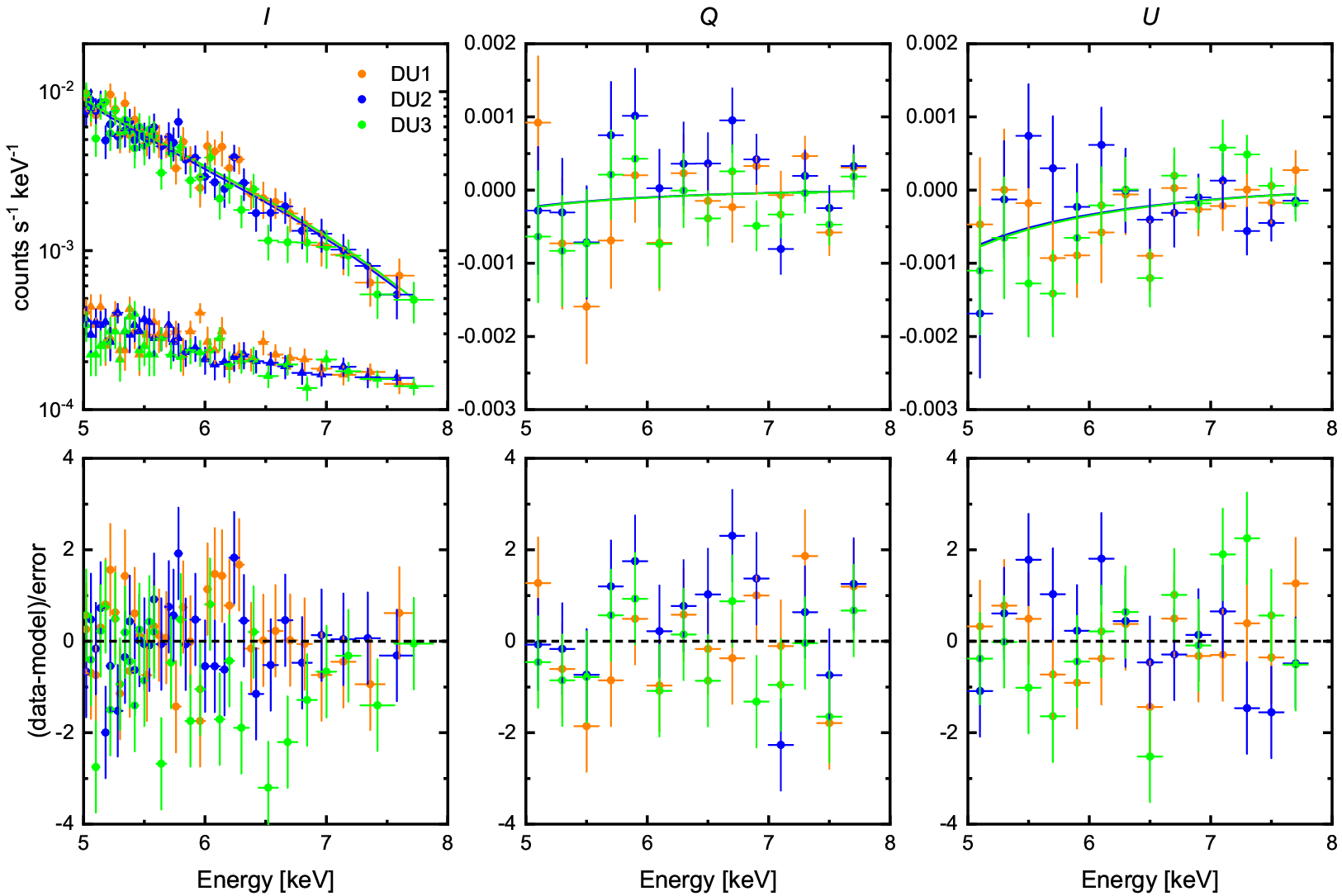}
    \caption{Spectropolarimetric fit results of the first {\it IXPE} observation data for Mrk 501, in the 2--5 keV band (top panel) and the 5--8 keV band (bottom panel), respectively. Panels present the fits to {\it IXPE} Stokes parameters $I$, $Q$ and $U$ with their associated residuals from left to right. In the $I$ fit panels, the triangles indicate the background count spectra for three DUs.}
    \label{IXPE_spec}
\end{figure}

\begin{figure}
    \centering
    \includegraphics[angle=0, scale=0.4]{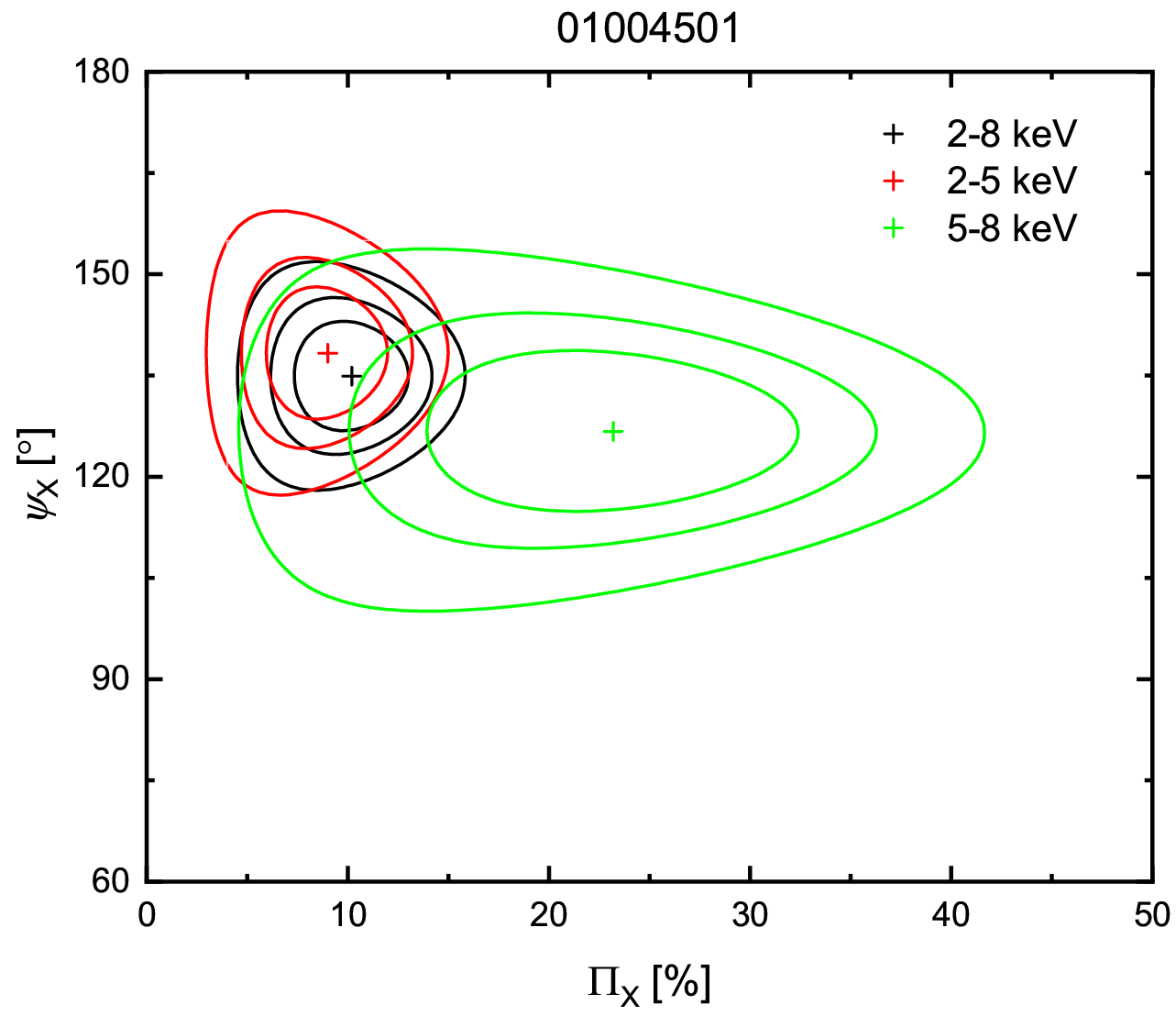}
    \includegraphics[angle=0, scale=0.4]{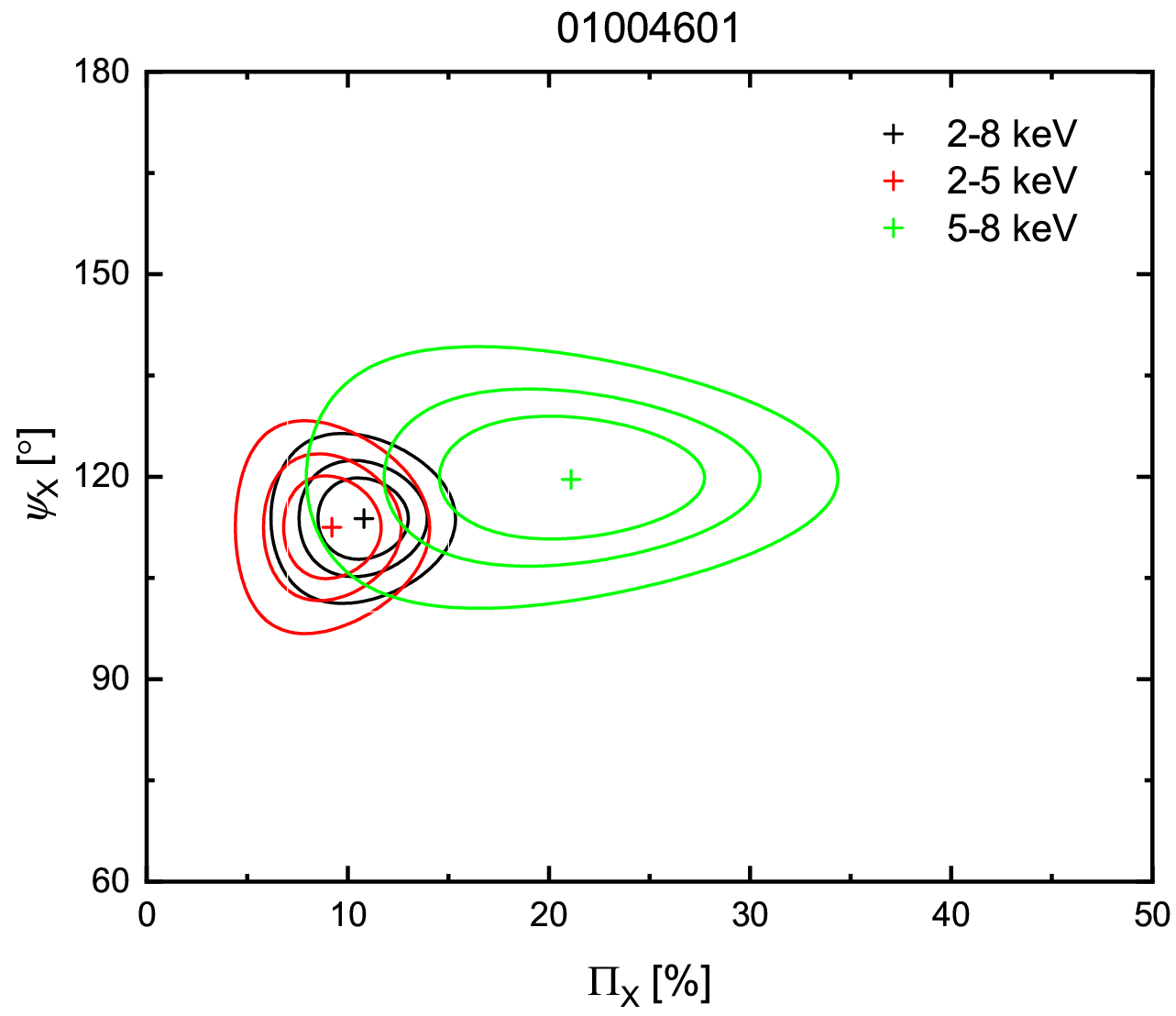}
    \caption{Confidence contours of X-ray polarization for the first (left panel) and second (right panel) {\it IXPE} observations in the 2--8 keV (black), 2--5 keV (red), and 5--8 keV (green) bands, which are derived through spectropolarimetric fits of the {\it IXPE} data. Contours are shown at confidence levels of 68\%, 90\% and 99\%, respectively.}
    \label{pol_contour}
\end{figure}

\begin{figure}
    \centering
    \includegraphics[angle=0, scale=0.5]{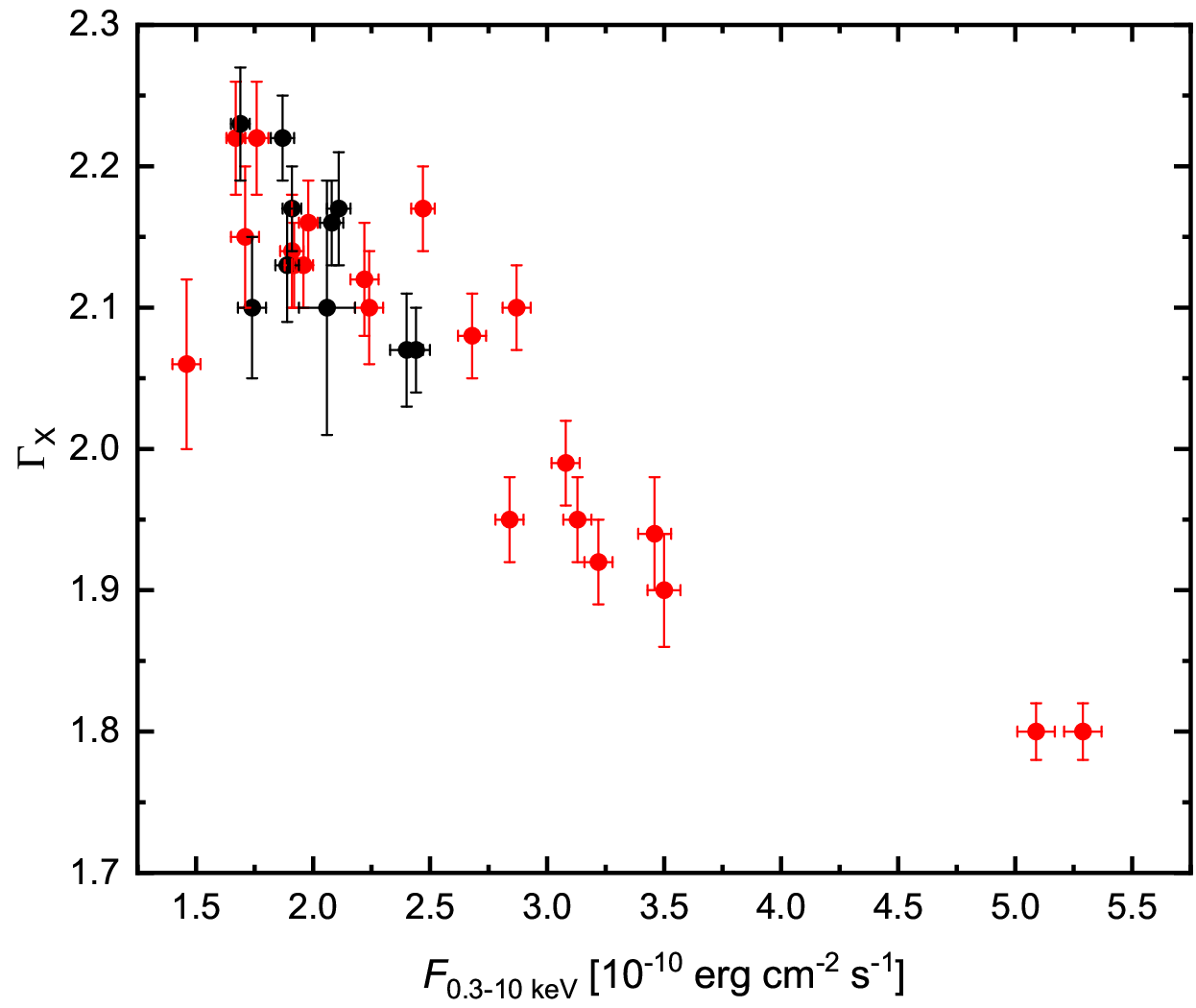}
    \caption{Photon spectral index ($\Gamma_{\rm X}$) vs. flux ($F_{\rm 0.3-10~keV}$) for Mrk 501, where $\Gamma_{\rm X}$ and $F_{\rm 0.3-10~keV}$ are derived with the {\it Swift}-XRT observations in the 0.3--10 keV band. The red points represent the data obtained simultaneously with the {\it IXPE} observation epochs, while the black points were derived from the {\it Swift}-XRT observations following the sixth {\it IXPE} observation, as listed in Table \ref{table_xrt}. The bootstrap method was employed to sample within the error range \citep[][]{MR0515681} and estimate the correlation coefficient ($r$) between $\Gamma_{\rm X}$ and $F_{\rm 0.3-10~keV}$, yielding a value of $r = -(0.85\pm0.03)$.}
    \label{Gamma_x-flux}
\end{figure}

\end{document}